\documentclass[a4paper,fleqn]{cas-sc}

\usepackage[numbers]{natbib}
\usepackage{longtable}
\usepackage{booktabs}

\def\tsc#1{\csdef{#1}{\textsc{\lowercase{#1}}\xspace}}
\tsc{WGM}
\tsc{QE}

\begin{document}
\let\WriteBookmarks\relax
\def\floatpagepagefraction{1}
\def\textpagefraction{.001}

\shorttitle{}    

\shortauthors{Batley and Saha}  

\title [mode = title]{Concurrent two-scale topology optimization of printable bone scaffolds via rectified-flow generation and exact latent-space refinement}  



%

\author[1]{Reza T Batley}[orcid=0009-0003-4863-626X]






\affiliation[1]{organization={Kevin T Crofton Department of Aerospace and Ocean Engineering, Virginia Tech},
            addressline={1600 Innovation Drive}, 
            city={Blacksburg},
            postcode={24060}, 
            state={VA},
            country={United States}}

\author[1]{Sourav Saha}[orcid=0000-0003-0237-0200]
\cormark[1]

\ead{souravsaha@vt.edu}



\cortext[1]{Corresponding author}



\begin{abstract}
Bone scaffolds must reconcile a macroscale mechanical response with microscale geometric feasibility, yet conventional approaches to their design restrict attainable constitutive properties.
This work presents a concurrent two-scale topology optimization framework; a conditional rectified flow generates periodic lattice microstructures, which are then refined in latent space by exact finite-element homogenization sensitivities.
The resulting unit-cell geometries and their spatial assignments are jointly optimized under mechanical and feature-size constraints.
In a distal rabbit femur benchmark, all eight flow-started refinements produce printable cells with maximum relative errors under 1\%  across six prescribed engineering moduli.
Constructing the implant from a single tiled cell reproduces intact whole-femur compliance within 0.1\% and regional strain energy within 0.35\% under four loading conditions.
In human femoral reconstruction, multiscale concurrent optimization produces a cortical mid-shaft scaffold with porosity 63\% while satisfying the compliance constraint within 2\% under walking and stair-climbing loads.
It achieves an 84\% porosity on a proximal cancellous scaffold with strain-energy deviations below 0.25\% under the same loads.
Analysis of the learned representations reveals distinct architectural families in latent space.
Further, the attention heads of the rectified-flow transformer specialize for spatial shifts -- learning the analogue of a finite-difference stencil -- and coordinate symmetries, with ablations showing their influence on generated topologies.
These results establish a route to printable, mechanically constrained scaffold designs. The framework is general enough for extending to many other multiscale structural design applications including topology-microstructure co-design in metal additive manufacturing. 
\end{abstract}


\maketitle

\section{Introduction}\label{sec:introduction}

The mechanical compatibility of an orthopedic implant with surrounding bone is central to its performance.
Too stiff, and it can redistribute load away from the host bone -- potentially causing stress shielding and bone resorption. Too compliant, and it can compromise structural stability \cite{Chmielewska2024,Safavi2023}.
Additively manufactured porous titanium provides a means of tuning effective stiffness -- subject to strength and bone ingrowth requirements \cite{Kelly2021}.
The feasibility of these scaffolds has been demonstrated experimentally in femoral defects \cite{VanderStok2012}, and clinically in the reconstruction of critical defects \cite{Liu2022a}.
Reducing effective stiffness alone, however, does not account for the anisotropy of bone and directional dependence of physiological loading \cite{Turner1990}.
Orthotropic lattices provide independent constitutive degrees of freedom for tailoring implant stiffness to the anisotropic mechanical requirements of bone.
For a transversely isotropic cancellous target, contemporary cell-design methods miss one modulus by 59--100\%.

Realizing these properties requires identifying geometrically admissible microstructures with prescribed constitutive responses.
This inverse design problem is non-unique: many distinct topologies may produce similar constitutive responses. 
Conditional generative models thus appear to offer a natural complement to numerical optimization in this domain.
By learning a distribution over geometries conditioned on the desired response, these models can represent this one-to-many inverse relationship and rapidly generate multiple candidate microstructures \cite{Wang2020}.
Numerical inverse homogenization approaches the same problem from the opposite direction: refining a particular geometry via sensitivities of its constitutive response \cite{Sigmund1994}.
Their complementary strengths suggest a division of labor in which generation identifies a suitable topology, and numerical homogenization refines its constitutive properties.
This is especially valuable when both constitutive accuracy and printability conditions must be satisfied. 
Local refinement can correct the mechanical response of a suitable topology, but cannot be relied upon to recover a printable geometry from an unsuitable initialization.
On a cancellous bone target, the best of 256 printable generated cells misses the prescribed moduli by 12.5\%, while exact refinement of the same cells closes to within 0.8--1.0\% from every generated start.
Alas, the very same refinement initialized from random latents fails to improve on its 31--41\% starting error within 400 steps under the thickness floor.
The role of generation is thus not merely in accelerating inverse homogenization, but supplying a geometrically admissible topology for subsequent refinement.

A suitably chosen latent representation provides a natural means of marrying these approaches.
By restricting optimization from full voxel space to a lower-dimensional manifold of plausible topologies, exact refinement need not directly manipulate every degree of freedom \cite{Zheng2023}. 
Existing methods have coupled these representations to learned constitutive predictors, typically neural networks, and use their gradients to guide latent optimization \cite{Seo2022,AbuMualla2025}.
Surrogate gradients, however, are not guaranteed to agree with those of the underlying homogenization problem, especially as optimization moves further from the training distribution \cite{Hoang2025,Trabucco2021}.
Refining the same starts with gradients from a learned property predictor reduces the \emph{predicted} error to below 0.03\%, while the exact error of the decoded cells remains 12--13\% -- essentially unchanged from initialization.
Exact adjoint refinement from the same starts reaches 0.8--1.0\%.
Differentiable numerical solvers offer an alternative, retaining the computed response of the decoded geometry as the optimization objective \cite{Hoyer2019,Xie2026}.
This preserves the distinction between the learned representation -- which constrains the geometric search space -- and the numerical model, which determines the mechanical response.

However, constitutive accuracy alone does not determine whether a generated microstructure can be realized.
Manufacturing imposes its own geometric constraints, particularly on thin structures and walls, whose dimensions also govern volume fraction and effective stiffness. 
Minimum length-scale constraints are well established in topology optimization \cite{Guest2004,Zhou2015}, and generative models have increasingly incorporated manufacturing considerations \cite{Jadhav2024}. 
Conditioning on stiffness and volume fraction does not independently control the distribution of local feature thickness, of course, and this distinction becomes crucial when unit cells are assigned physical dimensions.
At different cell scales, a fixed normalized thickness corresponds to different manufacturable feature sizes. 
Normalized feature thickness is therefore treated as an explicit design variable alongside the constitutive tensor and volume fraction. 
In particular, this work uses fifth-percentile thickness as representative rather than the minimum, which is sensitive to isolated voxel-scale features, to obtain a more robust measure of the thin regions present in each cell. 
Specifically, for a solid domain $\Omega_s$ with local-thickness field $t(x)$, 
\begin{align}
\label{eq:thickness}
    \mathcal{T}=\frac1L\,Q_{0.05}\left\{t(x) \mid x\in\Omega_s\right\}.
\end{align}
Here, $L$ is the physical unit-cell length (with the same units as $t$), $Q_{0.05}$ is the empirical fifth percentile via linear interpolation between ordered samples, and $t(x)$ is the diameter of the largest ball that contains $x$ and lies entirely within $\Omega_s$, as \citet{Hildebrand1997}.
The corresponding physical thickness is thus $\mathcal{T}\cdot L$, allowing a prescribed manufacturing threshold, $T_\mathrm{tol}$, to be expressed as $\mathcal{T}\geq \frac{T_\mathrm{tol}}{L}$.

This work combines these ideas in a conditional latent rectified flow framework for three-dimensional lattice generation. 
A generative model is trained to sample microstructures conditioned on the nine independent components of an orthotropic stiffness tensor $C^H$, volume fraction $v_f$ and the normalized fifth-percentile thickness $\mathcal{T}$. 
The generated geometry is refined by latent-space optimization via discrete finite-element-homogenization adjoints. 
Discrete sensitivities are propagated through the decoder to refine the three Young's and three shear moduli, and candidate geometries are screened against the prescribed thickness threshold. 
The resulting refinement is local by intent, only correcting constitutive responses, having outsourced the global-scale exploration of candidate topologies to the generative model.
This framework is evaluated through two-scale design of Ti-6Al-4V femoral implants, where optimized orthotropic stiffnesses must be realized by geometrically feasible lattice microstructures. 
Cortical and cancellous defect configurations provide contrasting mechanical targets under walking and stair-climbing loads.

The contributions of this work are as follows:
\begin{itemize}
    \item A conditional rectified-flow model that generates printability-aware unit cells from the full orthotropic stiffness tensor and volume fraction;
    \item Exact adjoint-homogenization refinement in latent space to drive cell properties to sub-1\% accuracy from every rectified-flow start while satisfying a feature size constraint; neither random initializations (30\%) nor a learned surrogate (12\%) suffices;
    \item Self-improvement via expert iteration: refined cells, re-encoded and exactly relabeled, retrain the generator raising the one-shot rate of printable cells within 5\% of bone targets from 4.0\% to 15.4\%;
    \item A two-scale constitutive design framework realized by online generation, refinement and exact constitutive feedback;
    \item Application to Ti-6Al-4V femoral implants, for both cortical and cancellous defects, under physiological walking and stair-climbing loads.
\end{itemize}

\section{Related work}\label{sec:related}

\subsection{Homogenization and multiscale constitutive design}

Early homogenization-based topology optimization replaced binary solid--void layouts with a continuous field of effective anisotropic porous materials \cite{Bendse1988}.
Inverse homogenization followed, seeking a periodic unit cell given a desired set of effective constitutive properties \cite{Sigmund1994}.
Density-based methods subsequently provided a simpler material-interpolation framework \cite{Bendse1999}, together with compact finite-element implementations for compliance minimization \cite{Sigmund2001,Andreassen2010}, and projection-based regularization for controlling minimum length scales \cite{Guest2004}. 
In making the material itself a design variable, macro- and microscales are coupled through repeated homogenization \cite{Coelho2007,Xia2014,Sivapuram2016}.
Conventional concurrent formulations require repeated macroscale equilibrium solutions and microscale constitutive evaluations as the design evolves.
Maintaining this coupling efficiently while retaining the sensitivities required for optimization remains a central challenge in multiscale constitutive design.

\subsection{Unit-cell libraries and data-driven multiscale design}

Precomputed unit-cell libraries provide an alternative to repeated on-demand microscale optimization by associating realizable geometries with their homogenized constitutive properties.
Early approaches constructed families of compatible microstructures for spatially-varying elasticity \cite{Schumacher2015}, and subsequently constrained two-scale optimization to a precomputed library of attainable material properties \cite{Zhu2017}.
More recently, generative modeling and active learning have enabled large libraries of bicontinuous, boundary-compatible cells, with demonstrated application to femoral implant design \cite{Wang2026}.
Parameterized triply periodic minimal-surface (TPMS) architectures, including the gyroid and IWP, provide an alternative family of established sheet-based lattice geometries for porous scaffolds \cite{Bobbert2017,TorresSanchez2021}.
However, this framework restricts elasticity to cubic symmetry -- with three independent stiffness constants -- and realizes optimized properties through discrete library lookup.
Thus, while data-driven libraries provide an effective bridge between constitutive optimization and geometric realization, their fidelity remains dependent on the coverage and resolution of the attainable property space.

\subsection{Learned representations for microstructure design}

The forward homogenization operator
\begin{align}
    \mathcal{H}:\rho(x)\mapsto C^H
\end{align}
maps a high-dimensional geometry to a much lower-dimensional set of high-level properties. 
In this way, many geometrically distinct units might occupy similar regions of property space. Inverse homogenization already operates under this assumption of non-uniqueness \cite{Sigmund1994}. 
One way to restrict the inverse problem is to constrain the admissible geometry to a lower-dimensional family. Spinodoid microstructures, for example, admit a compact continuous parameterization that supports inverse design within a smooth design space \cite{Kumar2020}.
Learned representations, in effect, generalize this idea to richer geometric families.
Deep generative modeling has been used to construct continuous latent representations of metamaterial systems, supporting interpolation between geometries and the design of spatially heterogeneous structures \cite{Wang2020}.
Similarly, graph variational autoencoders have been used to construct continuous design spaces of three-dimensional truss metamaterials, with latent variables optimized to match prescribed mechanical properties \cite{Zheng2023}.

Latent representations have subsequently been incorporated into multiscale optimization. 
One example couples a generative microstructure representation to a learned predictor of homogenized properties for multiscale topology optimization \cite{Seo2022}. 
A related approach used a variational-autoencoder representation of unit cells for thermo-mechanical multiscale design \cite{Chu2024}.
Janus couples an autoencoder with a constitutive-property predictor -- a separable neural architecture \cite{Batley2026d,Batley2025} -- to enable gradient-based inversion of prescribed thermal properties \cite{Batley2026,Batley2026a}.
Continuous latent variables have also been used to interpolate between multiple microstructure classes while jointly optimizing the local material volume fraction \cite{Wang2022a}. 
These works establish that learned continuous representations can provide effective design coordinates for complex local architectures and can be embedded within gradient-based multiscale topology optimization.

A complementary strategy is to generate microstructures directly from prescribed mechanical targets. Learned models have been used for non-iterative prediction of topology-optimized metamaterials \cite{Kollmann2020}.
More recently, conditional diffusion models have been used to generate three-dimensional microstructures from prescribed mechanical properties -- including voxel-based lattices, implicit lattice representations and curved truss architectures \cite{Zheng2025,Jadhav2024,AbuMualla2025}.
Guided diffusion has also been used to generate microstructures conditioned on effective elastic properties and boundary compatibility for multiscale topology optimization \cite{Feng2025}. 
Conditional generation learns a conditional distribution from material properties to candidate geometries, enabling multiple distinct microstructures to be sampled for the same constitutive target.

\subsection{Learned constitutive surrogates and differentiable numerical physics}

Repeated numerical homogenization is a major cost in concurrent topology optimization, motivating learned approximations of the map between local design variables and constitutive properties.
Neural network surrogates of microscale elasticity have been incorporated into gradient-based multiscale topology optimization \cite{White2019}.
Subsequent approaches have replaced repeated finite-element homogenization of spinodoid microstructures with a neural constitutive map whose automatic derivatives supply sensitivities with respect to local microstructure parameters \cite{Zheng2021}.
Similarly, deep networks have been trained to predict both homogenized properties and their shape sensitivities for concurrent multiscale structural optimization \cite{Black2023}.

A related strategy uses stiffness predictors to guide the inverse design of microstructures. Random-Walk Microstructures trains a convolutional stiffness surrogate on numerically homogenized cells, which is subsequently differentiated during inverse design \cite{Silverman2025}. 
Surrogate-guided optimization has also been extended to fully anisotropic elastic tensors using parameterized Kelvin-cell geometries \cite{Fu2025}.
Recent work has coupled a learned lattice decoder to a graph-convolutional predictor of equivalent mechanical properties in concurrent multiscale optimization \cite{Wu2026}.

An alternative here is to retain the numerical physics and differentiate through it. 
Neural reparameterization has demonstrated that design coordinates can be altered while retaining the original physics-based objective \cite{Hoyer2019}.
Neural parameterizations have been integrated with finite-element-based topology optimization \cite{Chandrasekhar2021}, while differentiable finite-element frameworks enable direct computation of numerical sensitivities for inverse design \cite{Xue2023}. Automatic differentiation of FFT-based homogenization provides another route to constitutive sensitivities \cite{Pundir2025}.
Differentiable physics solvers have also been used to guide diffusion-based inverse design of metamaterials \cite{Xie2026}.
Neither differentiable finite-element analysis nor physics-based microstructure sensitivities are themselves new, but the distinction between differentiating a surrogate and the discretized governing problem is especially relevant in inverse optimization.
Unconstrained learned homogenization surrogates have been shown to produce mechanically spurious solutions under gradient-based inverse design \cite{Keshav2026}.
The point is thus: predictive accuracy on sampled data and reliability in optimization are not identical objectives.

\subsection{Geometric realizability}

Gradient-based topology optimization commonly operates on relaxed or filtered representations that are subsequently projected towards binary geometries. 
Volume-preserving projection methods provide one route for controlling this transition \cite{Li2015}. 
In multiscale design, realization introduces additional constraints as locally optimized cells must remain connected and mutually compatible. Connectivity constraints have been introduced explicitly \cite{Du2018}. 
More recent approaches construct boundary-identical microstructure families to enforce geometric compatibility \cite{Feng2025}, or quantify physical compatibility by comparing the response of assembled microstructures against their homogenized equivalents \cite{Dai2026}. Geometric realizability also depends on feature size, which can be altered by projection or discretization, making geometric verification of the realized binary microstructure necessary.

\section{Computational Methods}\label{sec:methods}

\begin{figure}
    \centering
    \includegraphics[width=\linewidth]{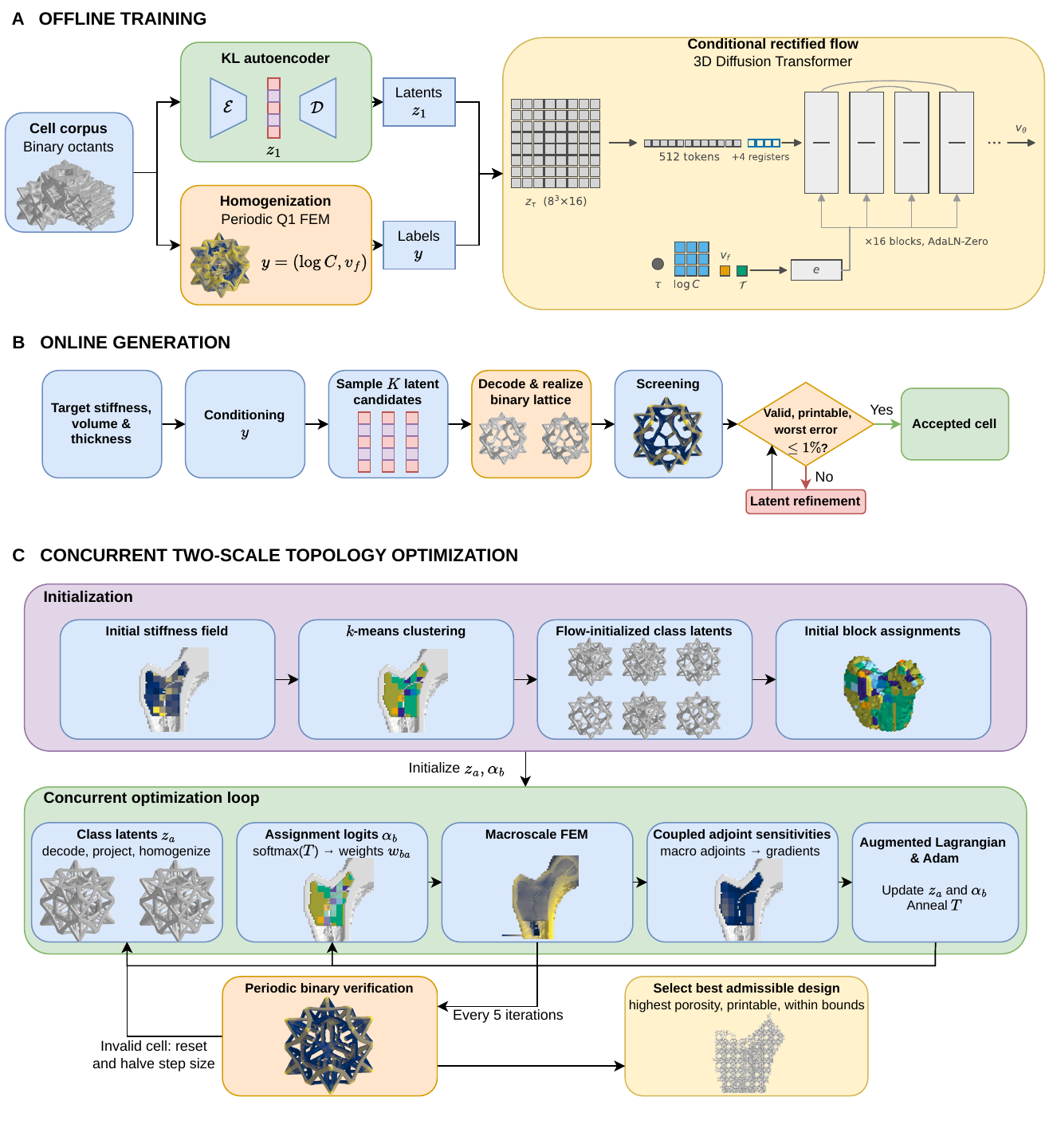}
    \caption{Overview of the proposed framework. a, Offline training of the conditional rectified-flow model. b, Online cell-level design. c, Concurrent two-scale topology optimization with coupled latent and assignment updates.}
    \label{fig:pipeline}
\end{figure}

The proposed framework proceeds in a latent space, combining initial microstructure generation with a conditional flow model, homogenization adjoint-based refinement, and gradient-based concurrent two-scale optimization (Figure \ref{fig:pipeline}).
During offline training, a $\beta$-variational autoencoder learns a latent representation of lattice geometries.
Periodic finite-element homogenization provides their constitutive properties.
A conditional rectified-flow model, parameterized by a diffusion transformer (DiT) \cite{Peebles2023}, learns to generate latent representations conditioned on those properties.
At inference, generated cells can be screened and refined by homogenization sensitivities for cell-level design.
For whole-structural optimization, flow-generated cells instead initialize a coupled problem in which latent microstructures and their spatial assignments are jointly optimized, their gradients chained to adjoints from homogenization and the frozen decoder.

\subsection{Unit-cell data and labels} \label{sec:labels}

Unit cells are periodic binary lattices on a $128^3$ voxel grid with mirror symmetry about the three mid-planes. They are stored and generated as a single $64^3$ octant, with the full cell recovered by mirroring. Because mirror symmetry makes each cell orthotropic in the cell frame, its homogenized elasticity is described by nine independent constants
\begin{align}
    c^H = \left(C_{11}^H,C_{22}^H,C_{33}^H,C_{12}^H,C_{13}^H,C_{23}^H,C_{44}^H,C_{55}^H,C_{66}^H\right).
\end{align}
The stiffness components are normalized by the reference solid modulus $E_0=1$MPa in constructing the conditional descriptors.
$\nu_s=0.3$ is the Poisson ratio of the reference solid. 

The dataset combines several families of cellular geometries. 
Cubic-symmetric cells are drawn from the public L-BOM library \cite{Wang2026} and relabeled with the Q1 finite-element solver used throughout this work. 
Parametric strut lattices are constructed from connected graphs of interior nodes and boundary anchors, with additional members introduced to increase topological diversity. 
These are further extended with elliptical plate elements and anisotropic variants, featuring preferentially aligned columns and plates. 
Further families include selectively thickened generated cells, low-volume derivatives obtained by thinning and stretching, and spinodoid architectures constructed from spatially correlated cosine fields.
Model-directed acquisition provides additional cells sampled from mechanically underrepresented regions -- including low-volume and strongly anisotropic targets. 
Although the L-BOM and isotropically thinned subsets retain cubic symmetry, the remaining families span orthotropic responses without imposing cubic elastic symmetry.  

Every cell is labeled with its homogenized stiffness tensor $C^H$, its volume fraction $v_f$ and its normalized fifth-percentile local thickness $\mathcal{T}$. The stiffness tensor is computed by periodic first-order (Q1) finite-element homogenization. This is a matrix-free solve of the six cell problems on the $128^3$ grid using an FFT-preconditioned conjugate-gradient method with void stiffness floor of $10^{-4}\cdot E_0$. 

Following the local-thickness definition of \cite{Hildebrand1997}, $t(x)$ denotes the diameter of the largest sphere containing $x$ and itself contained entirely within the solid domain $\Omega_s$. Then, the normalized thickness descriptor is as written in Equation \eqref{eq:thickness}.

The conditional generative model uses eleven material and geometric descriptors,
\begin{equation}
\begin{aligned}
y=\left[
\log\frac{C_{11}^H}{E_0},
\log\frac{C_{22}^H}{E_0},
\log\frac{C_{33}^H}{E_0},
\rho_{23},\rho_{13},\rho_{12},
\log\frac{C_{44}^H}{E_0},
\log\frac{C_{55}^H}{E_0},
\log\frac{C_{66}^H}{E_0},
v_f,\mathcal T
\right]^T\in\mathbb R^{11},
\end{aligned}
\end{equation}
with coupling ratios,
\begin{align}
    \rho_{ij}=\frac{C_{ij}^H}{\sqrt{C_{ii}^H C_{jj}^H}}, \quad ij\in\{(2,3),(1,3),(1,2)\}.
\end{align}
This parameterization preserves the nine independent stiffness components while accommodating the signed coupling terms. 
Each descriptor is standardized using the training-set means and standard deviations before embedding by the rectified-flow model.
Equivalently, the orthotropic response can be expressed through three Young's moduli, three shear moduli, and three Poisson ratios.
The inverse-design examples prescribe the six engineering moduli, leaving the Poisson ratios unconstrained. 

The pretraining set contains 117,048 distinct cells. Each is further included under its six axis permutations, with concomitantly permuted stiffness labels, yielding 702,288 training instances. Separate validation and test splits each contain 10,000 cells split by parent geometry group and stratified by family so that permuted instances of the same geometry cannot appear across dataset partitions.

Subsequent supervised fine-tuning (SFT) adds 2,689 finite-element refined cells. The validation and test sets remain unchanged for SFT.

\subsection{Latent autoencoder}\label{sec:vae}

\begin{figure}
    \centering
    \includegraphics[width=\linewidth]{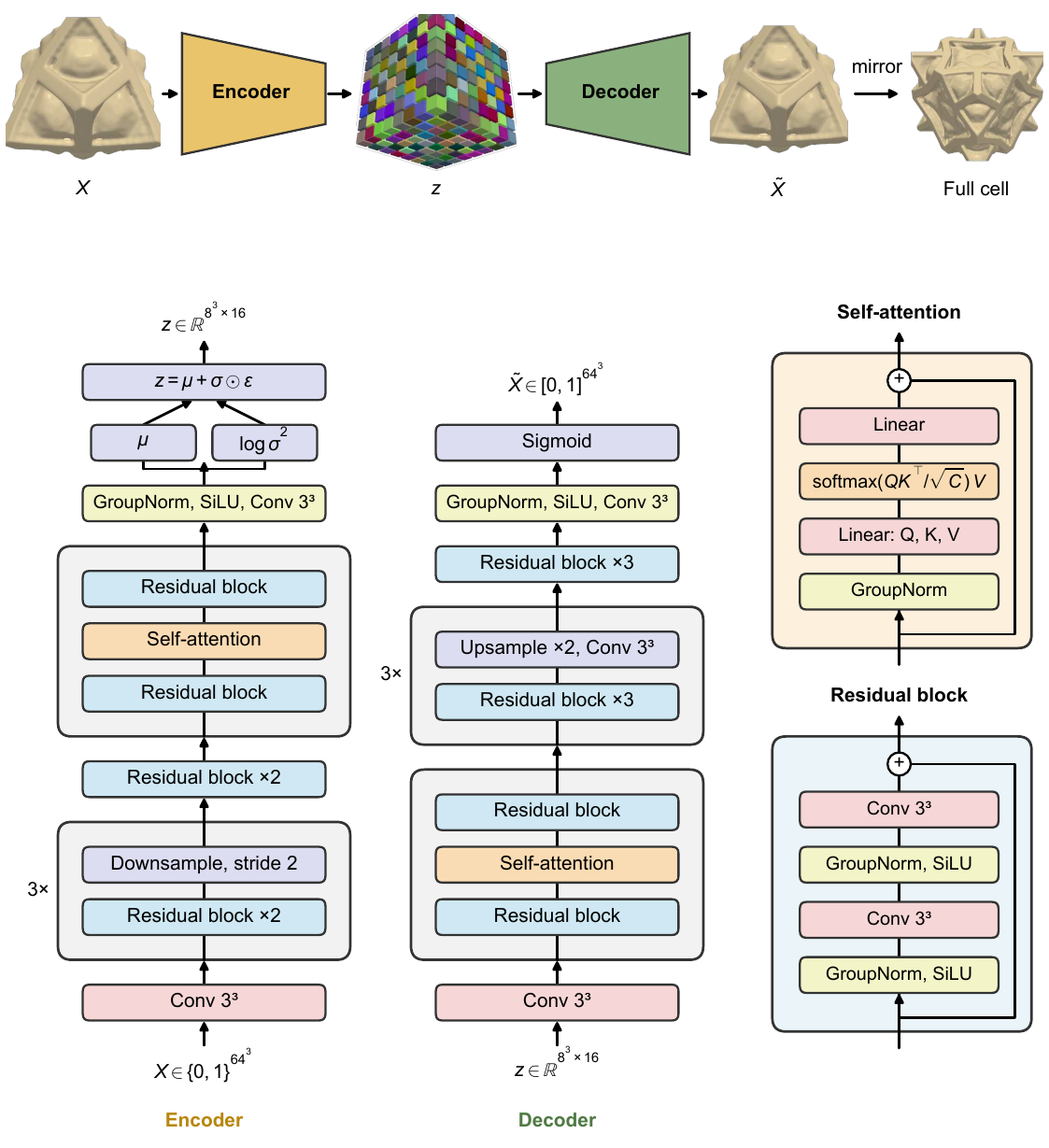}
    \caption{Autoencoder architecture and unit-cell reconstruction.}
    \label{fig:vae}
\end{figure}

A three-dimensional KL-regularized autoencoder -- following the latent-diffusion design \cite{Rombach2022} -- represents a binary symmetry octant $X\in\{0,1\}^{64^3}$ with a spatial latent $z\in\mathbb{R}^{8\times8\times8\times16}$. 
Let $d_z=8^3\cdot16=8,192$.
This reduces the number of scalar entries by a factor of 32 while maintaining a spatial latent grid.
The autoencoder is trained to reconstruct the geometry, without use of the constitutive-property labels.  
Beyond compression, the goal of the autoencoder is to supply a fixed representation of high-level features in which the conditional generative model is learned.
The encoder and decoder architectures, together with the reconstruction and mirroring operations are illustrated in Figure \ref{fig:vae}.

\paragraph{Probabilistic representation}

The autoencoder's probabilistic model factorizes as $p_\psi(X,z)=p_\psi(X\mid z)p_0(z)$, with standard Gaussian prior $p_0(z)\sim\mathcal{N}(0,I)$.
The encoder approximates its intractable posterior by a Gaussian,
\begin{align}
    q_\phi(z\mid X)&=\mathcal{N}\left(\mu_\phi(X),\operatorname{diag}(\sigma^2_\phi(X))\right), \\
    z&=\mu_\phi(X)+\sigma_\phi(X)\odot\epsilon,\\
    \epsilon&\sim\mathcal{N}(0,I).
\end{align}
In particular, the encoder outputs the log-variance $\ell_\phi=\log(\sigma^2_\phi(X))$ clipped to $[-30,20]$. 
$\odot$ denotes elementwise multiplication. 
The decoder outputs logits $a_\psi(z)$ and occupancy probabilities $d_\psi(z)=\operatorname{sigmoid}(a_\psi(z))$.
When frozen, this mapping is denoted $\mathcal{D}(z)=d_\psi(z)$.
Its reconstruction likelihood factorizes over the $N_X=64^3$ octant voxels,
\begin{align}
    p_\psi(X\mid z)=\prod_{e=1}^{N_X} d_{\psi,e}(z)^{X_e}(1-d_{\psi,e}(z))^{1-X_e}.
\end{align}

\paragraph{Encoder architecture} 

An input convolution is followed by four resolution levels with channel widths $32\cdot(1,2,4,8)$. 
Each level has two residual blocks (GroupNorm; SiLU; $3^3$ convolution twice; with a skip connection). 
The first three levels end with a stride-2 convolution reducing $64^3$ to $8^3$. 
At the $8^3$ bottleneck, a residual block, self-attention over the 512 spatial sites, and a second residual block completes the encoder backbone.
An output convolution gives the posterior mean $\mu_\phi$ and log-variance $\ell_\phi$.

\paragraph{Decoder} 

The decoder mirrors the encoder, with the middle stage first and three residual blocks per level, using nearest-neighbor upsampling followed by convolution, and outputs voxel logits.

\paragraph{Training objective}

The evidence lower bound (ELBO) satisfies,
\begin{align}
    \log p_\psi(X)&\geq \mathbb{E}_{q_\phi(z\mid X)}[\log p_\psi(X\mid z)]-D_\mathrm{KL}(q_\phi(z\mid X)\|p_0(z))
    \label{eq:elbo}.
\end{align}
The reconstruction term is the voxel-summed Bernoulli negative log-likelihood (NLL). 
The KL penalty is weighted, and the reconstruction term is augmented with latent perturbations. 
This yields a regularized reconstruction objective based on the two terms in Equation \eqref{eq:elbo}.
Namely,
\begin{align}
    \tilde z_b &=z_b+r_b\,\eta(s)\,s_\mu\,\zeta_b, \quad r_b\sim\mathcal U(0,1),\quad \zeta_b\sim\mathcal N(0,I),\\
    s_\mu^2&=\frac{1}{Bd_z}\sum_{b=1}^{B}\sum_{j=1}^{d_z} \big(\mu_{\phi,j}(X_b)-\bar\mu_{\mathcal B}\big)^2.
\end{align}
Here, $\bar \mu_\mathcal{B}$ is the mean over all posterior-mean entries in the minibatch, and $s_\mu$ is treated as constant during backpropagation. 
The added noise is scaled by the spread of posterior means, separately from the uncertainty in posterior sampling. 
At training step $s$, its maximum scale factor is
\begin{align}
    \eta(s)=\frac12\min\left\{\max\left\{\left(\frac{s-12000}{4000}\right),0 \right\},1\right\}.
\end{align}
The overall objective is
\begin{align}
 &\mathcal L(s)
 =\mathbb E\left[
 -\sum_{e=1}^{N_X}\left(X_e\log d_{\psi,e}(\tilde z)
 +(1-X_e)\log[1-d_{\psi,e}(\tilde z)]\right)
 +\lambda_{\rm KL}(s)\,D_\mathrm{KL}\big(q_\phi\,\|\,\mathcal N(0,I)\big)
 \right],\\
 &D_\mathrm{KL}(q_\phi\,\|\,\mathcal N(0,I))
 =\frac12\sum_{j=1}^{d_z}
 \left[\mu_{\phi,j}^2+\exp(\ell_{\phi,j})-1-\ell_{\phi,j}\right],\\
 &\lambda_{\rm KL}(s)
 =\exp\!\left[(1-f_s)\log 10^{-6}+f_s\log 10^{-3}\right],\quad f_s=\min\left\{\frac{s}{3000},1\right\}.
\end{align}
The expectation includes sampling of training cells, axis permutations, posterior noise and the additional latent perturbation. 
Reconstruction and KL terms are summed over their respective coordinates, and averaged over each minibatch. 
The autoencoder is trained for 24,000 steps with a minibatch size of 16, AdamW with no weight decay, a peak learning rate of $3\times 10^{-4}$, 1,000 warmup steps and cosine decay to 1\% of the peak rate.
Gradients are clipped to unit norm.

\paragraph{Flow-model latents}

After training, both the encoder and decoder are frozen.
The flow dataset uses the deterministic posterior mean of each oriented cell,
\begin{align}
    z_1(X)=\frac{\mu_\phi(X)-\mu_z}{\sigma_z}.
\end{align}
The flow operates in standardized latent coordinates, and thus sampled latents are transformed back to the decoder's native coordinates before decoding or optimization.
For $N$ oriented training rows with unstandardized latents $\hat z^{(n)}=\mu_\phi(X^{(n)})$, these scalar normalization constants are
\begin{align}
    \mu_z&=\frac{1}{Nd_z}\sum_{n=1}^{N}\sum_{j=1}^{d_z}
 \hat z_j^{(n)},\\
 \sigma_z^2&=\frac{1}{Nd_z}\sum_{n=1}^{N}\sum_{j=1}^{d_z}
 \left(\hat z_j^{(n)}-\mu_z\right)^2.
\end{align}

\subsection{Latent conditional rectified flow}

Inverse design requires a distribution of geometries compatible with a set of target descriptors, since multiple unit-cell topologies may plausibly share identical effective properties. 
In the frozen autoencoder representation, this translates to the problem of learning the conditional distribution $p_\mathrm{data}(z_1\mid c)$ of standardized geometry latents.
The conditioning descriptors, $j\in\{1,\dots,11\}$, are standardized componentwise via
\begin{align}
    \bar y_j=\frac{y_j-\mu_{y,j}}{\sigma_{y,j}}.
\end{align}
The condition is grouped into stiffness, volume and thickness. These have respective index sets $\mathcal{G}_C=\{1,\dots,9\};\mathcal{G}_v=\{10\}$ and $\mathcal{G}_T=\{11\}$.
The mask ${\mathcal{\widetilde M}}=[I_C,I_v,I_T]\in\{0,1\}^{3}$ specifies the supplied groups.
The corresponding component mask $\mathcal{M}\in\{0,1\}^{11}$ has entries $\mathcal{M}_j={\mathcal{\widetilde M}}_g$ for $j\in\mathcal{G}_g$, giving the availability condition $c=(\mathcal{\widetilde M},\mathcal{M}\odot\bar y)$.
For example, with volume omitted, $p_\mathrm{data}(z_1\mid c)$ conditions on stiffness and thickness while marginalizing over volume.
The conditional generative model is constructed as a probability transport from a Gaussian reference distribution to this data distribution, namely a rectified flow \cite{Lipman2022,Liu2022}.

\begin{figure}
    \centering
    \includegraphics[width=\linewidth]{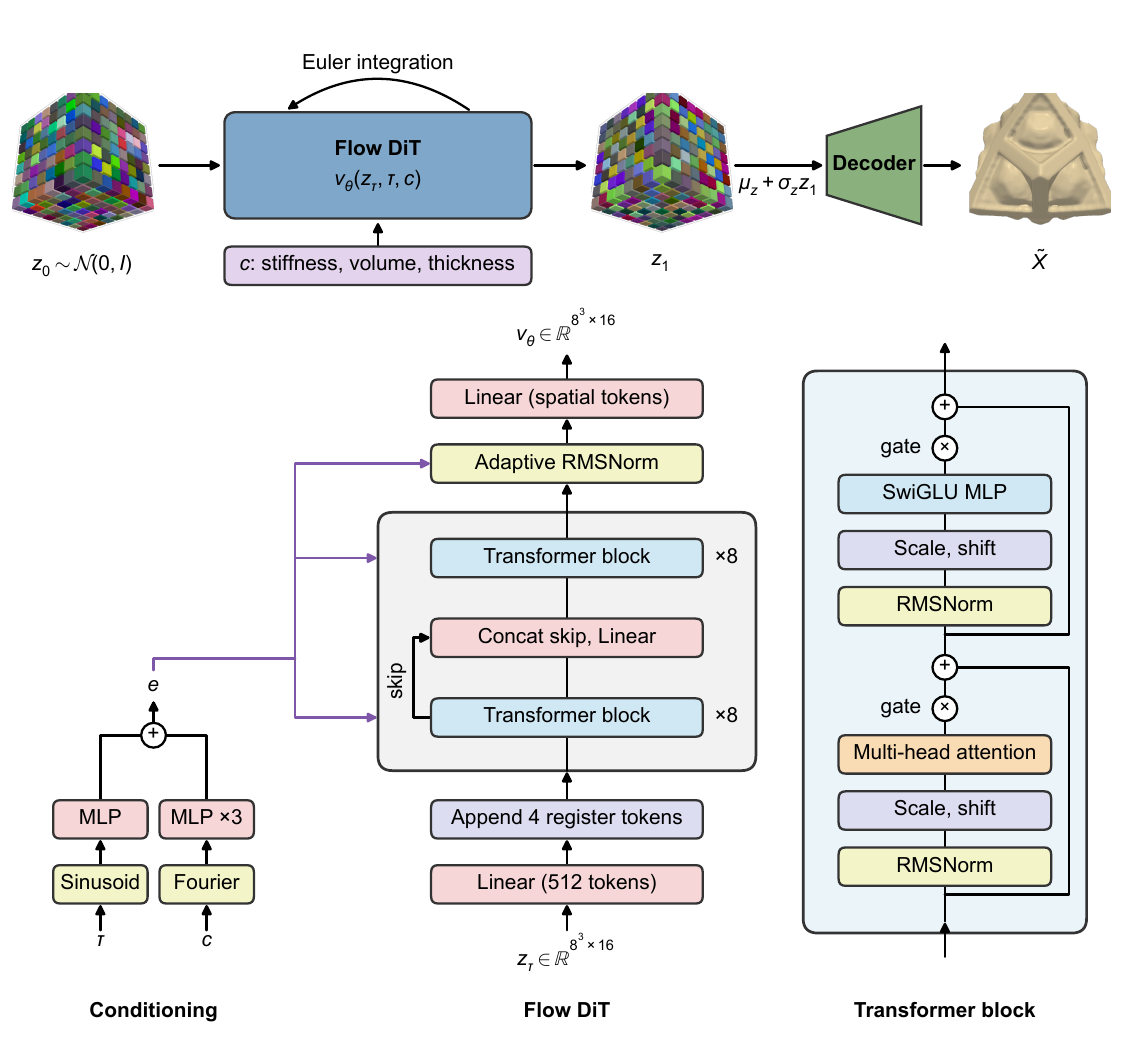}
    \caption{Conditional rectified-flow architecture and unit-cell generation.}
    \label{fig:dit}
\end{figure}

The conditional rectified-flow model and its diffusion-transformer architecture are shown in Figure \ref{fig:dit}.

\paragraph{Probability transport}

Let $p_\tau(z\mid c)$ be a path of latent distributions indexed by pseudo-time $\tau\in[0,1]$.
Furthermore, let this path have endpoints
\begin{align}
    p_{\tau=0}(z\mid c)&=\mathcal{N}(z;0,I),\\
    p_{\tau=1}(z\mid c)&=p_\mathrm{data}(z\mid c).
\end{align}
A velocity field $v(z,\tau,c)$ transports probability along this path, if it satisfies the continuity equation,
\begin{align}
     \frac{\partial p_\tau}{\partial\tau}(z\mid c)+\nabla_z\cdot[p_\tau(z\mid c)v(z,\tau,c)]=0.
\end{align}
Equivalently, under the usual regularity conditions for the flow, particles evolving according to
\begin{align}
    \frac{dZ_\tau}{d\tau}=v(Z_\tau,\tau,c),\qquad Z_0\sim\mathcal{N}(0,I),
\end{align}
have distribution $p_\tau(\cdot\mid c)$.

\paragraph{Straight-line probability path}

A straight-line path is specified by pairing, independently, a data latent $z_1$ with Gaussian noise $z_0\sim\mathcal{N}(0,I)$ to form the interpolant
\begin{align}
    z_\tau=(1-\tau)z_0+\tau z_1,
\end{align}
with constant pathwise velocity
\begin{align}
    \dot z_\tau=z_1-z_0.
\end{align}
For a fixed data endpoint, the intermediate distribution and a velocity that transports it are, for $\tau\in[0,1)$,
\begin{align}
    p_\tau(z\mid z_1)
 &=\mathcal N\left(z;\tau z_1,(1-\tau)^2I\right),\\
 u(z,\tau\mid z_1)
 &=\frac{z_1-z}{1-\tau}.
\end{align}
On a sampled path, this endpoint-conditioned velocity is simply
\begin{align}
     u(z_\tau,\tau\mid z_1)
 =\frac{z_1-((1-\tau)z_0+\tau z_1)}{1-\tau}
 =z_1-z_0.
\end{align}
The full conditional probability path, therefore, is the mixture
\begin{align}
p_\tau(z\mid c)=\int p_\tau(z\mid z_1)\,p_\mathrm{data}(dz_1\mid c).
\end{align}

Each endpoint-conditioned pair $(p_\tau(\cdot\mid z_1),u)$ satisfies the continuity equation. Averaging their probability fluxes gives the marginal transport velocity
\begin{align}
    v^*(z,\tau,c)&=\frac{\int u(z,\tau\mid z_1)p_\tau(z\mid z_1)p_\mathrm{data}(dz_1\mid c)}{p_\tau(z\mid c)}\\
    &=\mathbb E[z_1-z_0\mid z_\tau=z,c].
\label{eq:velocity}
\end{align}
This field satisfies the continuity equation for the mixture path. 
Note that while individual endpoint-paired training paths are straight, trajectories of the marginal field may curve: their velocities average over the possible endpoints compatible with the current latent and condition.

\paragraph{Conditional flow-matching objective}

Direct evaluation of Equation \eqref{eq:velocity} would require the conditional data distribution, which is unknown. 
Instead, a neural network is trained by regression on sampled endpoint pairs to act as a surrogate for the velocity field itself.
Let $\mathcal{P}$ denote the empirical distribution of training pairs $(z_1,y)$. 
Drawing a pair from $\mathcal{P}$, independent noise $z_0\sim\mathcal N(0,I)$, a time $\tau\sim\mathcal{U}(0,1)$ and a condition mask $\mathcal{\widetilde M}$, the objective is written
\begin{align}
    \mathcal L(\theta)=\mathbb{E}_{(z_1,y),z_0,\tau,\mathcal{\widetilde M}}\left[\frac{1}{d_z}\left\|v_\theta(z_\tau,\tau,c)-(z_1-z_0)\right\|_2^2\right].
\end{align}
Condition-mask sampling is independent of training geometry and descriptors.

\paragraph{Velocity network}

\begin{figure}
    \centering
    \includegraphics[width=\linewidth]{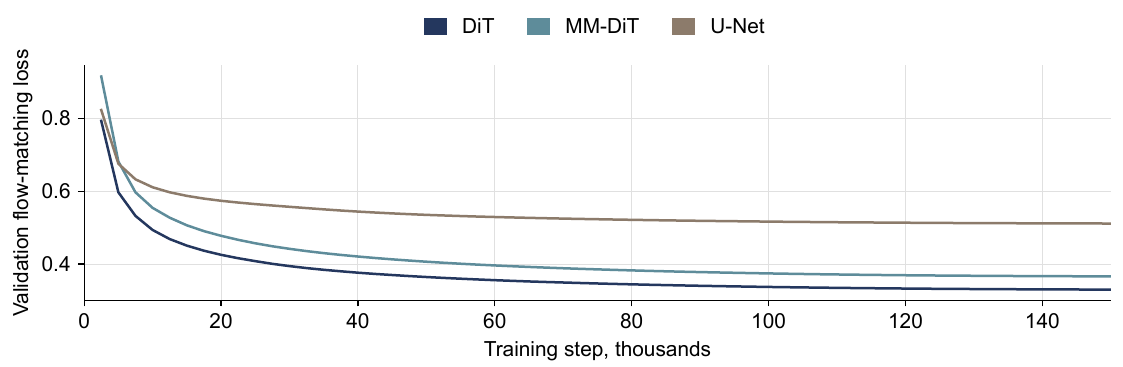}
    \caption{Validation flow-matching loss curves of the DiT, MM-DiT and U-Net backbones.}
    \label{fig:ablation}
\end{figure}

The flow velocity network, $v_\theta$, is a diffusion transformer (DiT) \cite{Peebles2023}, chosen following an ablation against a parameter- and data-matched ADM-style U-Net \cite{Dhariwal2021} and multimodal diffusion transformer (MM-DiT) \cite{Esser2024}. 
The three backbones were trained on identical data and schedules, with similar parameter counts (Table \ref{tab:backbone}).
The DiT attained the lowest validation flow loss (Figure \ref{fig:ablation}) and matched the alternatives in generation success on 48 printable held-out targets.
Differences in success are within the resolution of the target set.
It is notable that on the bone targets, none of the models generated a cell within 5\% maximum relative error.
This is an example motivating both the use of expert refinement, as well as numerical refinement of generated cells.

\begin{table}
\centering
\caption{Backbone ablation. Success is the fraction of targets
for which at least one of 64 samples is printable.}
\label{tab:backbone}
\begin{tabular}{lcccc}
\hline
Backbone & Parameters & Val. flow loss & Success $\leq 5\%$ & Success $\leq 10\%$ \\
\hline
\textbf{DiT} & 161.6M & \textbf{0.330} & 73\% & 98\% \\
MM-DiT & 159.8M & 0.366 & 73\% & 94\% \\
U-Net & 171.0M & 0.511 & 69\% & 98\% \\
\hline
\end{tabular}
\end{table}

The DiT operates on the $8^3$ latent grid, with patch size 1 -- i.e. 512 tokens -- plus 4 register tokens, width 720, 16 blocks and 8 attention heads.
The register tokens are learned auxiliary tokens that participate in self-attention but have no associated spatial location and are discarded before the final velocity projection \cite{Darcet2023}.
Each block uses RMSNorm with zero-initialized adaptive modulation, QK-normalized attention with three-dimensional rotary positional embeddings, and a SwiGLU MLP, with U-ViT-style long skip connections \cite{Bao2022}. 
Time and target descriptors modulate the transformer through the groupwise conditioning mechanism described as follows.

\paragraph{Conditioning}

Each standardized descriptor receives a Fourier feature vector
\begin{align}
    F_j(a)=\left[a,\sin(2\pi\omega_{j}a),\cos(2\pi\omega_ja)\right]^T,\quad \omega_j\in\mathbb{R}^{32},
\end{align}
where frequencies are initialized from a standard Gaussian and optimized during training. 
Fourier features provide a nonlinear representation of the continuous mechanical descriptors before groupwise embedding.
The features are concatenated within each group and passed through a separate two-layer SiLU MLP, denoted $\operatorname{MLP}_g$.
Its embedding is
\begin{align}
    e_g\left(\bar y,\mathcal{\widetilde M}_g\right)=\mathcal{\widetilde M}_g\operatorname{MLP}_g(\operatorname{concat}_{j\in\mathcal{G}_g} F_j(\bar y_j))+(1-\mathcal{\widetilde M}_g)n_g.
\end{align}
Here $n_g$ is the learned null embedding for a group $g$.
The shared conditioning embedding is
\begin{align}
    e\left(\tau,\bar y,\mathcal{\widetilde M}\right)=\operatorname{MLP}_t(\operatorname{SPE}(\tau))+\sum_{g\in\{C,v,T\}}e_g\left(\bar y, \mathcal{\widetilde M}_g\right).
\end{align}
Here, SPE is a sinusoidal positional embedding \cite{Vaswani2017}. 
The resulting embedding supplies the adaptive modulation parameters in each transformer block and the final output layer.

The shared conditioning embedding $e$ is passed through SiLU and a learned linear projection in order to obtain separate scale, shift and gate vectors for the attention and feedforward branches of each transformer block.
For an unnormalized token representation $h$, the adaptive modulation takes the form
\begin{align}
    \operatorname{AdaRMS}(h;e)=(1+s(e))\odot\operatorname{RMSNorm}(h)+b(e),
\end{align}
with the corresponding residual branch multiplied by a learned gate $g(e)$. 
Modulation parameters are shared across tokens within each sample. The linear projections producing them are initialized at zero in order that residual branches begin inactive.
Queries and keys are RMS-normalized before applying three-dimensional rotary positional embeddings (RoPE) \cite{Su2021}.
Because self-attention alone does not encode token positions, three-dimensional RoPE introduces spatial positional information along the three latent axes into the attention operation.
This is applied to spatial tokens only.
A final adaptively modulated normalization and zero-initialized linear projection recover the $8\times8\times8\times16$ latent velocity field from the spatial tokens.

\paragraph{Guidance} 

For classifier-free guidance \cite{Ho2022}, all three condition groups are replaced by their learned null embeddings with probability 0.1 during training. 
Independently, the volume group is also replaced by its null embedding with probability 0.3, so that targets can be generated without prescribing $v_f$.
Complete conditioning, conditioning on stiffness and thickness alone, and fully unconditioned training occurs with probabilities 0.63, 0.27 and 0.1 respectively.
At sampling time, the guided velocity is
\begin{align}
    v^\mathrm{CFG}_\tau=v_\varnothing+w\,(v_c-v_\varnothing),
    \label{eq:cfg}
\end{align}
with guidance scale $w$, and where $v_c=v_\theta(z_\tau,\tau,c)$ and $ v_\varnothing=v_\theta(z_\tau,\tau,\varnothing)$. Here, $\varnothing$ denotes the fully masked condition $\mathcal{\widetilde M}=(0,0,0)$. 

\paragraph{Training}

The optimizer is AdamW, with batch size 256, weight decay 0.01, cosine learning-rate decay, and an exponential moving average of the weights used for all evaluation. 
The base model is trained for 150,000 steps at learning rate $10^{-4}$ (2,000 warm-up steps, EMA decay 0.9999). 
Each expert-iteration round fine-tunes the base model for 10,000 steps on the enlarged dataset at learning rate $5\times10^{-5}$ (250 warm-up steps, EMA decay 0.999), upweighting the most recently acquired cells eightfold.

\subsection{Inverse design of unit cells}\label{sec:inverse}

\paragraph{Generation}

The rectified-flow model defines a continuous transport from Gaussian noise $z_0\sim\mathcal{N}(0,I)$ to a sample from the learned conditional latent distribution  $p_\theta(z\mid c)$. 
In the following examples, the volume-fraction conditioning group is masked. 
A sample is drawn by integrating the transport ODE,
\begin{align}
    \frac{dz_\tau}{d\tau}=v_\theta(z_\tau,\tau,c),\quad z_{\tau=0}\sim\mathcal{N}(0,I),\quad \tau\in[0,1].
\end{align}
At inference, to improve adherence to property targets, classifier-free guidance (CFG) modifies the learned velocity field as Equation \eqref{eq:cfg} with guidance scale $w=4$ .
The resulting transport ODE is integrated with 16 forward-Euler steps from $z^{(0)}=z_0$ via 
\begin{align}
    z^{(n+1)}=z^{(n)}+h\cdot v^\mathrm{CFG}\left(z^{(n)},nh,c\right),~h=\frac{1}{16},~n\in\{0,\dots,15\}
\end{align}
The resulting generated latent $z_1=z^{(16)}$ is subsequently decoded to $\tilde X=\mathcal{D}(\mu_z+\sigma_z z_1)\in[0,1]^{64^3}$ -- the occupancy probabilities on the symmetry octant.

\paragraph{Realization}

To the decoded unit cells, a realization function $\mathcal{R}:[0,1]^{64^3}\to\{0,1\}^{128^3}$ is applied.
This proceeds as follows:
\begin{enumerate}
    \item thresholding at 0.5: deleting voxels with an occupancy probability under a half;
    \item overwriting the prescribed connective-boundary mask voxels with their fixed solid/void values;
    \item mirroring to a full $128^3$ cell; and
    \item removing solid islands not connected to the largest 6-connected component.
\end{enumerate}
The result is checked to form a single connected solid that spans the cell in all three directions, and that the boundary mask remains intact after solid islands are removed. Candidates that fail are discarded.

\paragraph{Screening}

In this work, for each target, 64 such cells are generated and realized. They are then evaluated by loose-PCG-tolerance ($10^{-2}$; found to change the worst-modulus error by at most 0.06\%) Q1 finite-element homogenization. Feasibility -- i.e. the realized geometry is valid with $\mathcal{T}\geq \frac{T_\mathrm{tol}}{L}$ -- is assessed. Feasible candidates are ranked by their maximum relative error over the six engineering moduli,
\begin{align}
    \epsilon_{\max}
=\max_k\left|\frac{e_k}{e_k^\star}-1\right|,
\quad
e=(E_1,E_2,E_3,G_{23},G_{13},G_{12}).
\end{align}
These are obtained from the homogenized compliance,
\begin{align}
    S^H=(C^H)^{-1},\quad E_i=\frac{1}{S^H_{ii}},\quad G_{23}=C^H_{44},\quad G_{13}=C^H_{55}, \quad
    G_{12}=C^H_{66},
    \label{eq:moduli}
\end{align}
using engineering-shear Voigt notation.
If the best feasible candidate satisfies $\epsilon_{\max}\leq\epsilon_\mathrm{tol}$, it is accepted directly.
In this work the acceptance tolerance is taken to be $\epsilon_\mathrm{tol}=0.01$.
Otherwise, the eight best feasible candidates are refined.

\paragraph{Refinement}

Each selected latent is optimized using Adam for up to 80 steps with learning rate $0.02\sigma_z$. 
The objective is the squared logarithmic discrepancy between target and computed engineering moduli:
\begin{align}
    \mathcal{L}(z)=\frac16\sum_{k=1}^{6}\left(\log\frac{\hat e_k(z)}{e_k^\star}\right)^2.
\end{align}
Given $\tilde X=\mathcal{D}(z)$, let $\tilde{X}_\xi\in[0,1]$ denote the occupancy predicted by the frozen decoder at voxel $\xi$. 
For differentiable homogenization, projection occurs via a smoothed Heaviside,
\begin{align}
    m_\xi&=\rho_{\min}+H_\beta(\tilde{X}_\xi),\\
    H_\beta(s)&=\frac{\tanh\left(\frac{\beta}{2}\right)+\tanh\left(\beta\left(s-\frac12\right)\right)}{2\tanh\left(\frac{\beta}{2}\right)},
\end{align}
which continuously approximates binarization.
Here $\beta$ is taken to be 20 and $\rho_{\min}=10^{-4}$. 
The prescribed mask is held fixed, and the resulting field is mirrored to the full periodic cell. 
For homogenization, element stiffness is left to vary linearly with $m_\xi$ -- i.e. without a SIMP penalization exponent. 

The predicted moduli $\hat e$ are obtained via Q1 homogenization of $m=\{m_\xi\}_{\xi=1}^N$, which gives $C^H$, then conversion as Equation \ref{eq:moduli}.
For each imposed macroscopic strain case $i$, the periodic cell problem is solved for the displacement fluctuation $\chi^{(i)}$. 
By stationarity of these cell-problem solutions, the sensitivity of the homogenized stiffness with respect to each $m_\xi$ requires no separate adjoint solve. 
Let $A_{pq}^{(i)}$ be the imposed macroscopic strain and $\chi_q^{(i)}$ be the periodic displacement fluctuation for a load case $i$.
Indeed, defining the local strain field as
\begin{align}
    \varepsilon_{pq}^{(i)}=A^{(i)}_{pq}-\frac12\left(\partial_p\chi_q^{(i)}+\partial_q\chi_p^{(i)}\right),
\end{align}
the sensitivity with respect to each $m_\xi$ is
\begin{align}
    \frac{\partial C^H_{ij}}{\partial m_\xi}=\frac{E_0}{|\Omega_\mathrm{cell}|}\int_{\Omega_{\mathrm{cell},\xi}}\varepsilon_{pq}^{(i)}D^{(0)}_{pqrs}\varepsilon^{(j)}_{rs}~dx.
\end{align}
Here $i,j\in\{1,\dots,6\}$ index the macroscopic strain cases in Voigt order, $\Omega_\mathrm{cell}$ is the periodic cell domain, and $D^{(0)}$ is the dimensionless fourth-order elasticity tensor of the solid material.
The gradient is propagated through the chain $z\to \tilde X\to m\to C^H\to \hat e\to \mathcal{L}$, using automatic differentiation through the projection operator $H_\beta(\tilde X)$ and the decoder $\mathcal{D}(z)$.
The full chain from latent to objective is
\begin{align}
\frac{d\mathcal L}{dz}
=
\overbrace{\frac{\partial\mathcal L}{\partial\hat e}}^{\text{Evaluate loss}}
\underbrace{\frac{\partial\hat e}{\partial C^H}}_{\text{Convert moduli}}
\overbrace{\frac{\partial C^H}{\partial m}}^{\text{Homogenize}}
\underbrace{\frac{\partial m}{\partial \tilde X}}_{\text{Project}}
\overbrace{\frac{\partial \tilde X}{\partial z}}^{\text{Decode}}.
\end{align}
The annotations denote precisely that operation which the gradient reverses.
At every other step, the latent is decoded and realized as a binary cell and re-evaluated by Q1 homogenization (relative PCG residual $10^{-4}$) and the thickness criterion.
Over the eight starts, the best feasible iterate is retained, and a start halts once it reaches $\epsilon_\mathrm{tol}$.

\subsection{Macroscale analysis}

The structural response is evaluated by a two-scale finite-element model under quasistatic, small-strain linear elasticity. 
The macroscopic domain $\Omega$ comprises retained host bone $\Omega_\mathrm{bone}$ and a designated replacement region $\Omega_d$.
Each design block is assigned the homogenized orthotropic stiffness of a periodic unit cell.
Displacement continuity across the bone--implant interface models perfect bonding.

\paragraph{Constitutive representation}

Letting $b(\xi)$ denote the design block containing macroscopic finite element $\xi$, one may write the constitutive tensor as
\begin{align}
    C_\xi=
    \begin{cases}
        C_{b(\xi)}^{\mathrm{implant}},&\rm{if} \quad \Omega_\xi\subseteq\Omega_d,\\
        C_\xi^{\mathrm{bone}},&\rm{otherwise}.
    \end{cases}
\end{align}
The constitutive tensor of the implant is constructed from the nine independent components of orthotropic elasticity, written in Voigt notation, ordered $(xx,yy,zz,yz,xz,xy)$, as
\begin{align}
    D_{b}^\mathrm{implant}=\begin{pmatrix}
        C_{11}&C_{12}&C_{13}&0&0&0\\
        C_{12}&C_{22}&C_{23}&0&0&0\\
        C_{13}&C_{23}&C_{33}&0&0&0\\
        0&0&0&C_{44}&0&0\\
        0&0&0&0&C_{55}&0\\
        0&0&0&0&0&C_{66}
    \end{pmatrix}_b.
\end{align}
$D^\mathrm{implant}_b$ is the matrix representation of $C^\mathrm{implant}_b$ in Voigt notation.
In the concurrent formulation, $C_b$ denotes this tensor before binarization.
The cell axes are aligned with the axes of the macroscopic grid, which are themselves aligned with the anatomical Heller frame \cite{Heller2005}.

Homogenization of each unit cell uses a reference modulus $E_0$ and Poisson ratio $\nu_s=0.3$. For an implant with solid modulus $E_s$, the stiffness supplied to the macroscale solver is
\begin{align}
    C_{b}^\mathrm{implant}=\frac{E_s}{E_0}C_{b}^H,
\end{align}
where $C_b^H$ contains the nine independently retained orthotropic constants obtained by homogenization.
A small ersatz stiffness, with relative floor $\rho_{\min}=10^{-4}$, regularizes the nominally void phase in each cell homogenization. 
The modulus rescaling follows from the linear dependence of the cell stiffness operator on the reference solid modulus. 

\paragraph{Finite-element equilibrium}

For a load case $i$, the macroscopic displacement $u^{(i)}$, strain $\varepsilon^{(i)}$ and stress $\sigma^{(i)}$ satisfy
\begin{align}
    \varepsilon_{pq}(u^{(i)})&=\frac12\left(\partial_q u_p^{(i)}+\partial_p u_q^{(i)}\right), \\
    \sigma_{pq}^{(i)}&=C_{pqrs}\varepsilon_{rs}(u^{(i)}),\\
    \partial_q\sigma_{pq}^{(i)}&= 0\quad\textrm{in }\Omega,
\end{align}
subject to prescribed displacement constraints and external forces.
Bone material assignments, geometry, boundary and loading conditions are specified for each design problem in Section \ref{sec:results}.
For virtual displacements $w$ and external force functional $\ell^{(i)}$, the weak form is thus
\begin{align}
    \int_\Omega \varepsilon_{pq}(w)C_{pqrs}\varepsilon_{rs}(u^{(i)})~dx=\ell^{(i)}(w).
\end{align}

The domain is discretized using trilinear eight-node hexahedral elements with $2\times2\times2$ Gauss integration.
The elemental and assembled stiffness operators are
\begin{align}
    K_\xi&=\int_{\Omega_\xi} B_{\xi}^TD_\xi B_\xi~dx,\\
    K(x)u^{(i)}&=f^{(i)}.
\end{align}
Here, $B_\xi$ is the strain--displacement matrix and $D_\xi$ is the assigned constitutive matrix. $x$ collects the nine constitutive coordinates of each implant design block with their dependence on the latent and assignment variables introduced in Section \ref{sec:concurrent}. 
Constrained degrees of freedom are eliminated from the equilibrium system.
The remaining equations are solved using matrix-free, Jacobi preconditioned conjugate gradients in double precision with a relative target residual of $10^{-8}$ checked every 50 iterations.

\paragraph{Mechanical responses}

Mechanical constraints are expressed relative to the corresponding intact-bone reference and evaluated using the same mesh, loading and boundary conditions.
Two structural responses are considered.
For whole-domain compliance,
\begin{align}
    J^{(i)}=f^{{(i)}T}u^{(i)}=u^{(i)T} K u^{(i)}.
\end{align}
For elastic strain energy, within the replacement region $\Omega_d$,
\begin{align}
    J^{(i)}&=\frac12 u^{(i)T}K_d u^{(i)},\\
    K_d&=\sum_{\xi:\Omega_\xi\subseteq\Omega_d}K_\xi.
\end{align}
The regional response measures load transfer through the implant, and is used as a proxy for the local mechanical stimulus relevant to stress shielding.
Unlike whole-domain compliance, it depends explicitly on both the displacement field and the constitutive properties within $\Omega_d$.

For either response, define
\begin{align}
    r^{(i)}(x)=\frac{J^{(i)}(x)}{J^{(i)}_0}-1,
\end{align}
where $J^{(i)}_0$ is the response of the intact bone reference under the same load case.
Thus, positive and negative deviations represent response greater and smaller than the intact value, respectively.

\paragraph{Adjoint sensitivities}

The macroscale optimization requires derivatives of the mechanical responses with respect to the constitutive design coordinates $x_{b,k}$. 
These coordinates denote component $k$ of the physical block stiffness, with $k=1,\dots,9$ following the component ordering defined in Section \ref{sec:labels}.
$b$ indexes the design blocks.
Differentiating the weak form with respect to $x_{b,k}$ gives
\begin{align}
    \int_\Omega \varepsilon_{pq}(w)C_{pqrs}\varepsilon_{rs}\left(\frac{\partial u^{(i)}}{\partial x_{b,k}}\right)~dx=-\int_{\Omega_b}\varepsilon_{pq}(w)\frac{\partial C_{pqrs}}{\partial x_{b,k}}\varepsilon_{rs}(u^{(i)})~dx.
\end{align}
Here $\Omega_b$ is the region occupied by design block $b$.
The derivative of the constitutive tensor disappears outside the block, of course.

For a mechanical response $J^{(i)}$, introduce the adjoint displacement $\lambda^{(i)}_p$ defined by
\begin{align}
    \int_\Omega \varepsilon_{pq}(\lambda^{(i)})C_{pqrs}\varepsilon_{rs}(w)~dx=\frac{\partial J^{(i)}}{\partial u^{(i)}}[w],
\end{align}
for all admissible $w$.
The right-hand side denotes the directional derivative with respect to displacement.
This gives the total sensitivity,
\begin{align}
    \frac{d J^{(i)}}{d x_{b,k}}=\left.\frac{\partial J^{(i)}}{\partial x_{b,k}}\right|_u-\int_{\Omega_b} \varepsilon_{pq}(\lambda^{(i)})\frac{\partial C_{pqrs}}{\partial x_{b,k}}\varepsilon_{rs}^{(i)}~dx.
\end{align}

\paragraph{Whole-domain compliance}

For compliance,
\begin{align}
    J^{(i)}=\int_\Omega \varepsilon_{pq}^{(i)}C_{pqrs}\varepsilon_{rs}^{(i)}~dx.
\end{align}
Taking $J^{(i)}=f^{(i)T}u^{(i)}$, its displacement derivative equals the forcing functional with no dependence on design coordinates.
Consequently, the adjoint is the equilibrium displacement itself $\lambda^{(i)}=u^{(i)}$.
Thus,
\begin{align}
    \frac{dJ^{(i)}}{d x_{b,k}}=-\int_{\Omega_b}\varepsilon_{pq}^{(i)}\frac{\partial C_{pqrs}}{\partial x_{b,k}}\varepsilon_{rs}^{(i)}~dx.
\end{align}

\paragraph{Regional strain energy}

Within the implant region $\Omega_d$,
\begin{align}
    J^{(i)}=\frac12\int_{\Omega_d}\varepsilon_{pq}^{(i)}C_{pqrs}\varepsilon_{rs}^{(i)}~dx.
\end{align}
Its displacement derivative gives
\begin{align}
    \int_{\Omega}\varepsilon_{pq}(\lambda^{(i)})C_{pqrs}\varepsilon_{rs}(w)~dx=\int_{\Omega_d}\varepsilon_{pq}(w)C_{pqrs}\varepsilon_{rs}^{(i)}~dx.
\end{align}
This requires an additional adjoint solve.
The regional-energy sensitivity is
\begin{align}
    \frac{d J^{(i)}}{dx_{b,k}}=\frac12\int_{\Omega_b}\varepsilon_{pq}^{(i)}\frac{\partial C_{pqrs}}{\partial x_{b,k}}\varepsilon_{rs}^{(i)}~dx-\int_{\Omega_b}\varepsilon_{pq}(\lambda^{(i)})\frac{\partial C_{pqrs}}{\partial x_{b,k}}\varepsilon_{rs}^{(i)}~dx.
\end{align}
The first integral accounts for the explicit change in strain energy within the replacement region while the second accounts for displacement redistribution through the structure.
Finally, for the relative mechanical response
\begin{align}
    r^{(i)}=\frac{J^{(i)}}{J^{(i)}_0}-1,
    \label{eq:deviations}
\end{align}
the sensitivity is
\begin{align}
    \frac{dr^{(i)}}{d x_{b,k}}=\frac{1}{J^{(i)}_0}\frac{d J^{(i)}}{d x_{b,k}}.
\end{align}
These supply the mechanical constraint gradients for concurrent optimization.

\subsection{Concurrent two-scale design}
\label{sec:concurrent}

The concurrent formulation optimizes the latent representation of unit cells and their spatial assignments under macroscale mechanical constraints.
At each iteration, the macroscale response is evaluated using the homogenized tensors of the current decoded microstructures and its sensitivities are propagated both to the cell latents and their assignments.

\paragraph{Initialization}

An initial distribution of orthotropic constitutive properties is obtained by blockwise optimization.
Each design block is parameterized by the nine transformed stiffness coordinates spanning the printable library $\mathbb L$.
The mean solid fraction is estimated via local linear model fitted to the nearest 32 library cells in the coordinate space.
The initial problem minimizes the mean predicted solid fraction subject to the macroscale response constraints $|r^{(i)}|\leq\delta$ with $\delta=0.02$, using the method of moving asymptotes (MMA) \cite{Svanberg1987}.
A move limit of 0.1 is applied.
After each update, the block coordinates are projected to within $\rho_\mathrm{lib}=0.03$ of their nearest library point.
The solve comprises 80 iterations.

Block tensors are parametrized by the logarithms of the nine orthotropic constants, bounded by the range of the printable training cells with positive constants.
The continuum design therefore excludes cells with negative couplings (0.7\% of the training set).
The resulting block tensors are clustered into $N_c=12$ classes by $k$-means on the standardized logarithms of the stiffness constants.
Each class target is the componentwise geometric mean of its member-block stiffness constants.
Flow model generation and geometric screening, as described in Section \ref{sec:inverse}, provide the initial latent microstructure for these classes.
For each class, the initial latent is the valid, printable screened sample with the smallest worst-case error relative to the class target.
A class without such a sample starts from the latent of the class with the nearest diagonal stiffness.
This preliminary optimization determines the initial class distribution and microstructures, but subsequent optimization acts directly on decoded cells.

\paragraph{Design variables}

Each class $a\in\{1,\dots,N_c\}$ is represented by a latent variable $z_a$. Its decoded unit cell is
\begin{align}
    \tilde X_a=\mathcal{D}(z_a),
\end{align}
from which the projected density $H_\beta(\tilde X_a)$, homogenized orthotropic tensor $C_a^H(z_a)$ and solid fraction
\begin{align}
    v_{f,a}(z_a)=\left\langle H_\beta(\tilde X_a)\right\rangle
\end{align}
are evaluated using the procedures of Section \ref{sec:inverse}.

The spatial assignment of classes to design blocks is also optimized.
Each block $b$ carries logits $\alpha_b\in \mathbb R^{N_c}$ which are converted to class weights via
\begin{align}
    w_{ba}=\frac{e^{\alpha_{ba}/T}}{\sum_{j=1}^{N_c}e^{\alpha_{bj}/T}}.
\end{align}

The block constitutive tensor and solid fraction are thus
\begin{align}
    C_b&=\frac{E_s}{E_0}\sum_{a=1}^{N_c} w_{ba}C_a^H(z_a),\\
    v_b&=\sum_{a=1}^{N_c} w_{ba}v_{f,a}(z_a).
\end{align}
Here $C_b$ is the implant-scaled block stiffness, while $v_b$ is the relaxed volume fraction.

This continuous relaxation permits gradient-based optimization of the class distribution \cite{Stegmann2005}. The logits are initialized
\begin{align}
    \alpha_{ba}=4\mathbb{1}[a=a_0(b)],
\end{align}
where $a_0(b)$ is the initial cluster assignment.
The softmax temperature is reduced geometrically from $T_0=0.5$ to $T_1=0.02$ over the optimization to encourage increasingly discrete assignments.
Classes may become unused as the assignment hardens.

\paragraph{Optimization problem}

The objective is the mean solid fraction volume across the design blocks.
Let $N_b$ denote the number of design blocks and $N_L$ the number of prescribed load cases. 
Then,
\begin{align}
    &\min_{\{z_a\},\{\alpha_b\}}\left\{\frac{1}{N_b}\sum_{b=1}^{N_b} v_b \right\},\\
    &\textrm{subject to}\quad |r^{(i)}|\leq\delta,\quad i=1,\dots,N_L;\quad\tau_a\leq\tau_\mathrm{tol},\quad a=1,\dots,N_c.
\end{align}
In this work, $\delta=0.02$ and $\tau_\mathrm{tol}=0.04$ are chosen.

The geometric constraint uses a differentiable approximation to the fraction of solid material lying in features thinner than the prescribed threshold.
Specifically, $\tau_a$ is the fraction of projected solid removed by a soft morphological opening constructed from erosion and dilation using alternating cube and cross max-pooling operations.
The structuring-element radius is selected from the prescribed thickness threshold in voxel units, following the length-scale-control principle of morphological filtering \cite{Sigmund2007}.
The constraint $\tau_a\leq \tau_\mathrm{tol}$ provides a margin relative to the 5\% thin-feature fraction associated with the fifth-percentile thickness condition.
The exact thickness of the final geometry is evaluated separately.

\paragraph{Sensitivities and optimization}

The macroscale adjoints provide the derivatives of each response deviation with respect to the block constitutive tensors.
Differentiating the block constitutive interpolation gives the gradients with respect to each class tensor
\begin{align}
    \frac{\partial r^{(i)}}{\partial C_a^H}=\frac{E_s}{E_0}\sum_{b=1}^{N_b} w_{ba}\frac{\partial r^{(i)}}{\partial C_b}.
\end{align}
These gradients are propagated to the class latents through the discrete Q1 homogenization sensitivities and decoder chain rule developed in Section \ref{sec:inverse}.
Componentwise,
\begin{align}
    \frac{d r^{(i)}}{d z_a}=\frac{E_s}{E_0}\sum_{b=1}^{N_b}\sum_{k=1}^9 w_{ba}\frac{\partial r^{(i)}}{\partial x_{b,k}}\frac{\partial c_{a,k}^H}{\partial z_a}.
\end{align}
Here $x_{b,k}$ is the physical stiffness component of block $b$ and $c_{a,k}^H$ is the corresponding homogenized component of class $a$, evaluated at $E_0$.

Derivatives with respect to assignment logits follow by differentiating the softmax weights.
The volume objective and morphological constraint are differentiated through the density projection and decoder.

The constrained problem is treated using an augmented Lagrangian with penalty parameter 2, and multiplier updates at every iteration.
Latents and assignment logits are updated by Adam with learning rates $0.02\sigma_z$ and 0.1, respectively.
The concurrent optimization is run for at most 100 iterations, with the temperature schedule applied over the interval.

\paragraph{Verification and design selection}

Candidates are periodically evaluated in their discrete form, here every five iterations, and at every iteration for which the responses of the soft design satisfy
\begin{align}
    \max_i |r^{(i)}|\leq 2\delta.
\end{align}
Each block is assigned the class of maximum weight
\begin{align}
    a(b)=\arg\max_a w_{ba}.
\end{align}
Each class is realized by $\mathcal R$ as a binary unit cell.
Its nine orthotropic constants are recomputed by Q1 homogenization, initially to a relative residual of $10^{-4}$, and its thickness is evaluated.
Should the realized cell be invalid or fail geometric requirements, its latent is restored to the most recent printable state and update step size halved.

The verified constitutive tensors are assigned to the macroscale blocks and structural responses are computed without soft material mixing.
The corresponding mean block porosity is
\begin{align}
    \phi=1-\frac{1}{N_b}\sum_{b=1}^{N_b}v_{f,a(b)}.
\end{align}
An iterate is admissible if every used class produces a valid, printable binary cell and the resulting macroscale responses satisfy $|r^{(i)}|\leq\delta$ for each load case.
Failure of any used class renders the iterate inadmissible.

Among assembled, verified iterates, the reported design is that of the highest porosity.
The optimization terminates after 100 iterations, or earlier by patience stopping.
Patience stopping is triggered when the best admissible porosity has not improved by 0.1\% over 20 iterations.
The selected design is re-evaluated by Q1 homogenization at a relative residual of $10^{-8}$.

Consequently, the reported mechanical responses and porosities are evaluated on the binary realized microstructures.

\section{Analysis of Results}\label{sec:results}

\begin{figure}
    \centering
    \includegraphics[width=\linewidth]{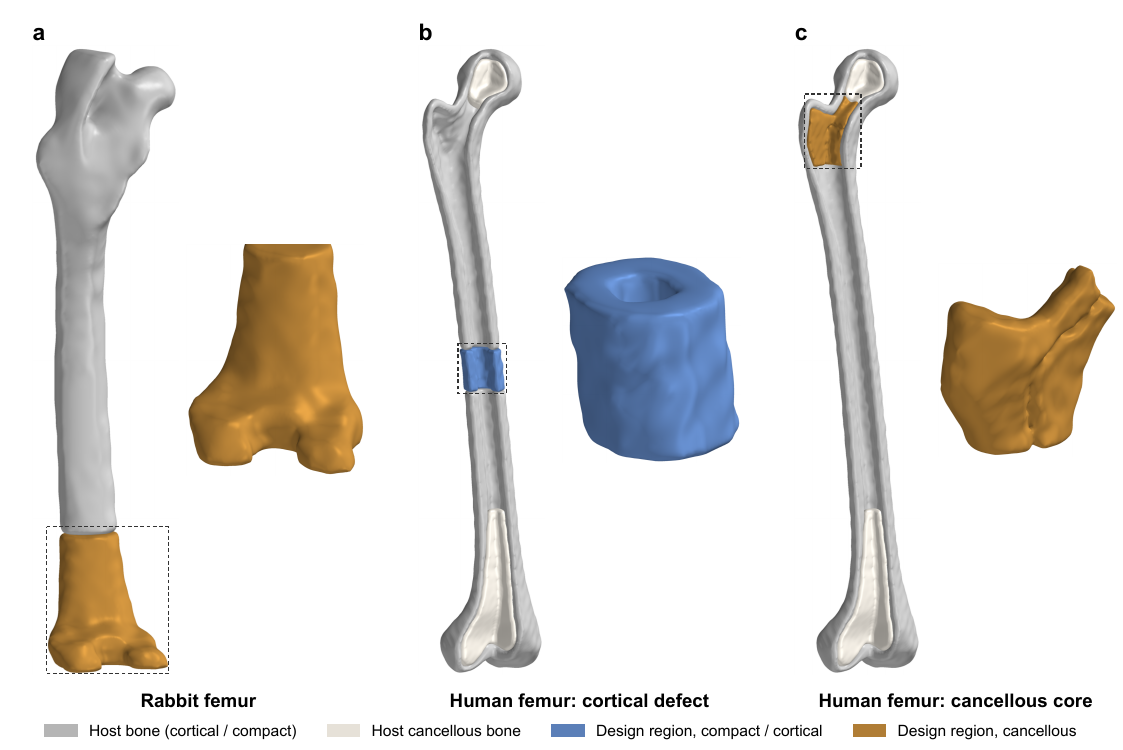}
    \caption{Anatomical configurations and design regions. a, The cancellous distal portion of a rabbit femur. b, Human femur with a cortical mid-shaft defect. c, Human femur with a proximal cancellous replacement.}
    \label{fig:implants}
\end{figure}

Three anatomical configurations are considered (Figure \ref{fig:implants}). 
The distal rabbit femur provides cancellous bone targets for evaluating cell-level inverse design and structural responses.
Two human femoral defect configurations are then used to assess concurrent two-scale optimization: a cortical mid-shaft, and proximal cancellous defects.
The following results examine geometric feasibility and constitutive matching, then evaluate the resulting implant designs under the prescribed mechanical loading conditions.

\subsection{Benchmark: unit-cell inverse design for a rabbit femur}

\begin{figure}
    \centering
    \includegraphics[width=\linewidth]{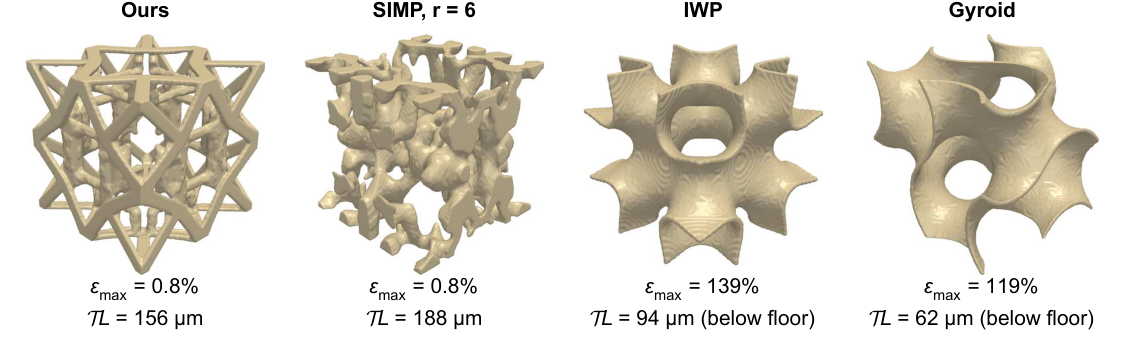}
    \caption{Representative realized unit cells obtained by the proposed method and baseline approaches. Maximum relative engineering-modulus error $\epsilon_{\max}$ and physical feature size $\mathcal T L$ are indicated.}
    \label{fig:r1gallery}
\end{figure}

The rabbit femur benchmark uses a distal implant region of 7,267 voxels within an 80.5mm femur, discretized at 0.5mm. 
The geometry is from the ANU Skullbook dataset \cite{Carro2020}. 
The cancellous target is transversely isotropic with longitudinal Young's modulus of 2,500MPa, longitudinal-to-transverse modulus ratio of 2 and prescribed shear moduli.
The femur is clamped at its distal face, and loaded at the proximal end by unit axial, transverse (two directions) and torsional loads applied separately.
Note that as the model is linear and all responses are reported relative to the intact femur, load magnitudes do not affect the comparison.
The anisotropy ratios and Poisson ratios are modeling assumptions.
For flow conditioning, the three coupling ratios are taken from the nearest printable training cell in log-diagonal stiffness space, while refinement is evaluated purely against the six engineering moduli.
The implant is assembled from 4mm periodic titanium cells, subject to a feature thickness of 156$\mu$m.

The proposed method is compared with i) nearest-cell lookup from the cubic-symmetric L-BOM library, ii) I-graph-wrapped (IWP) and iii) gyroid sheet lattices, and iv) SIMP inverse homogenization. The TPMS cells -- IWP and gyroid -- are thickness-tuned to match the longitudinal modulus.
SIMP uses the same six-modulus objective with periodic Q1 sensitivities, density filtering and projection continuation, with four random initializations at filter radius 3 and eight at radius 6.

Figure \ref{fig:r1gallery} shows representative realized cells.
The proposed cell achieves a maximum relative error of 0.80\% over the six engineering moduli, with a physical feature size on-target at 156$\mu$m.
All eight flow-initialized refinement runs produced printable cells with errors between 0.80\% and 0.98\%.
At radius 3, SIMP matches the moduli (0.67--0.98\%) but only with features near 94$\mu$m, below the floor.
Radius 6 yields printable features (188--250$\mu$m) in all eight runs, but only two reach the 1\% tolerance: realized errors are 0.80--1.95\% in seven runs, and one exceeds 30\% after removal of disconnected material.

The two components do not, individually, suffice.
Of 256 printable flow samples, the best misses the target by 12.5\%.
For this target, the samples occupy a single narrow design family (solid fraction 0.116--0.119), so screening alone cannot close the gap.
Refinement from random latents never yields a printable cell (31--41\%, 62$\mu$m features).
Refining the flow samples with gradients from a learned surrogate predictor drives the predicted error below 0.03\%, while the exact error remains at 12\%.
Only exact refinement from flow-generated cells reaches 0.80--0.98\% -- from every start.

\begin{figure}
    \centering
    \includegraphics[width=\linewidth]{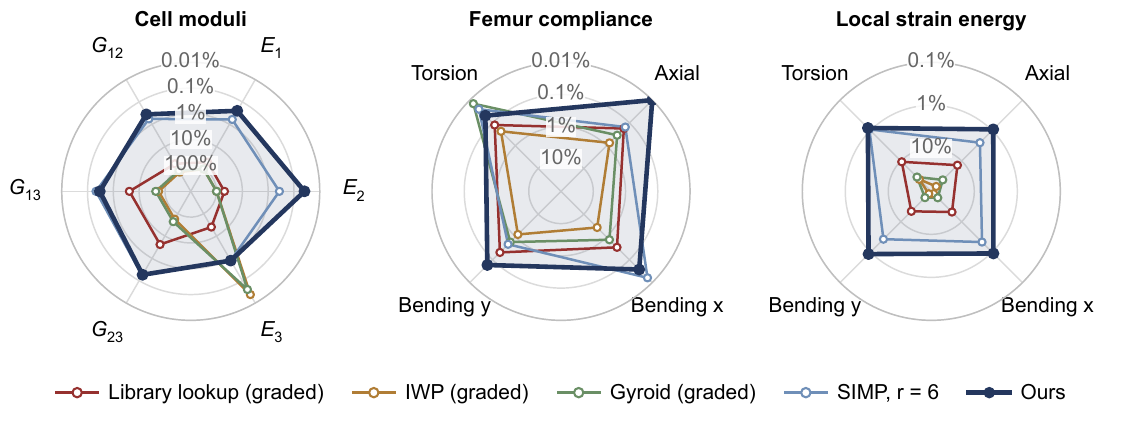}
    \caption{(\emph{left}) Relative errors in the six prescribed engineering moduli for the cancellous bone target; (\emph{center}) relative errors in the target whole-femur compliance under axial, two bending, and torsional loading; (\emph{right}) relative strain-energy deviations. Baselines in the left panel are uniform cells; the center and right show their graded implants; radial scales are logarithmic and reversed.}
    \label{fig:r1moduli}
\end{figure}

The componentwise errors are shown in Figure \ref{fig:r1moduli} (\emph{left}). Nearest-library lookup gives a maximum error of 59\%, while IWP and gyroid give 139\% and 119\%, respectively, although both TPMS cells match the longitudinal modulus.
In varying their thickness to match the longitudinal modulus, the resultant respective feature sizes of 94 and 62$\mu$m also violate the prescribed floor.
In contrast, the proposed cell matches all six prescribed moduli within 1\% while maintaining the desired feature thickness.

Flow-initialized refinement requires 75--105s per start, plus 184s of screening shared across both targets; SIMP at radius 6 requires 163--364s per run on the same hardware (an NVIDIA B200 GPU). 
The principal distinction is consistency: every flow-initialized start meets the 1\% tolerance, compared with two of eight SIMP runs.

The unit cells were subsequently assembled into distal femoral implants and evaluated under axial loading, two bending directions and torsion.
Figure \ref{fig:r1moduli} (\emph{center}) compares whole-femur compliance with the intact model.
The proposed implant reproduces compliance within 0.1\% under all four loads, with no greater than 0.35\% strain-energy deviations in the implant region.
SIMP also performs well, with compliance deviations within 0.5\% and regional energy deviations within 1.3\%.

Tiled uniformly (not shown), the library implant produces compliance deviations up to 11\%, while IWP and gyroid exhibit changes of approximately 1--5\%. Their implant-region strain-energy deviations reach 25\%, 37\% and 32\%, respectively.
Thus, relatively small global compliance changes do not necessarily imply agreement in the regional response of the implant.

Graded comparisons were also evaluated to account for the spatial variation permitted in L-BOM-style design.
These graded variants, shown in the center and right panels of Figure \ref{fig:r1moduli}, reproduce whole-femur compliance within 2.6\% and implant-region strain energy within 2.4\%.
Their elementwise strain-energy errors are 11--22\% for the graded library, 34--62\% for gyroid and 40--84\% for IWP, compared with 0.8--0.9\% for the proposed cell.
The graded library, gyroid and IWP designs require 16, 12 and 12 distinct cells, respectively, of which 9, 6 and 4 fall below the thickness floor.
These comparisons use independently implemented graded optimization procedures rather than the published L-BOM pipeline of \citet{Wang2026} -- and without access to its full lattice library, which is not public.

\begin{figure}
    \centering
    \includegraphics[width=\linewidth]{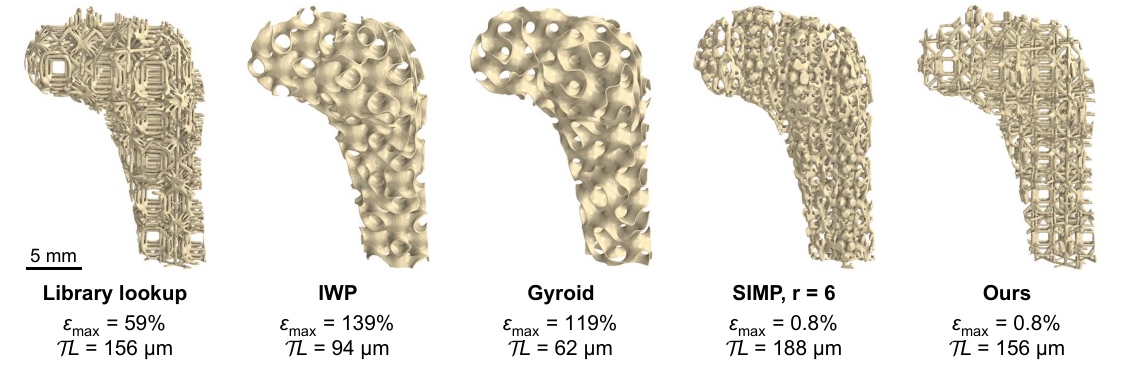}
    \caption{The realized distal rabbit femur implants constructed from the proposed and baseline lattice designs.}
    \label{fig:r1implant}
\end{figure}

\begin{figure}
    \centering
    \includegraphics[width=\linewidth]{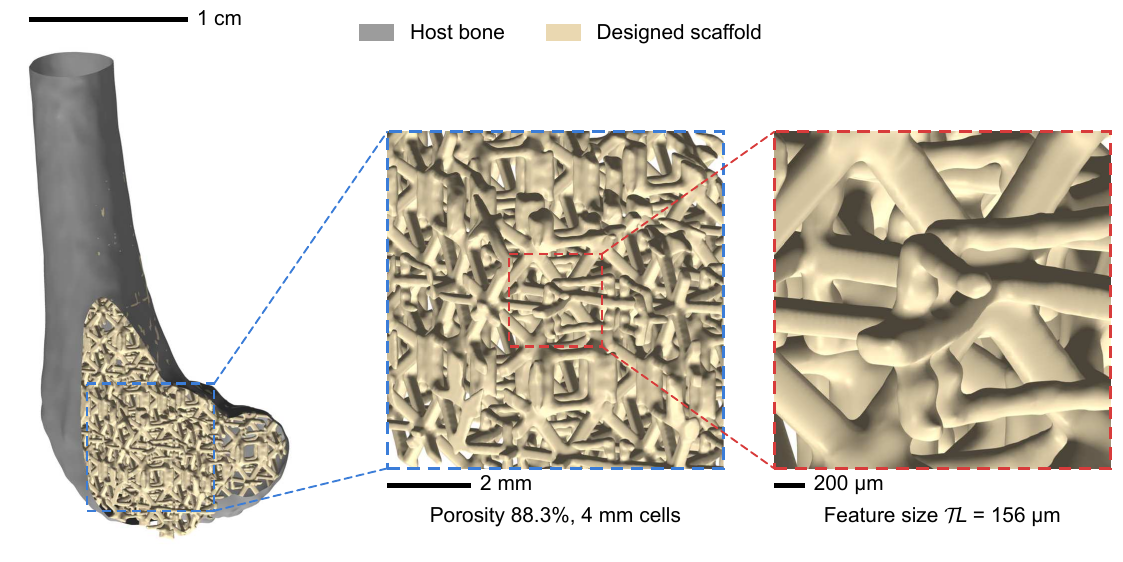}
    \caption{Multiscale visualization of the proposed distal rabbit femur scaffold.}
    \label{fig:r1design}
\end{figure}

The assembled implants themselves are shown in Figure \ref{fig:r1implant}. 
The proposed method achieves the prescribed six-modulus accuracy and close agreement both in global and local femur responses using a single printable cell -- no spatial grading required.
This is shown in greater detail, across scales, in Figure \ref{fig:r1design}.
Indeed, spatial grading of the baseline lattices recovers global stiffness but leaves local strain-energy deviations of up to 84\%, whereas a single cell matched to all six moduli recovers both.
This distinction matters in orthopedic reconstruction, where bone adaptation is governed by local mechanical stimuli -- such as strain-energy density -- rather than whole-structure stiffness alone \cite{Huiskes2000,Turner1998}.

\begin{figure}
    \centering
    \includegraphics[width=\linewidth]{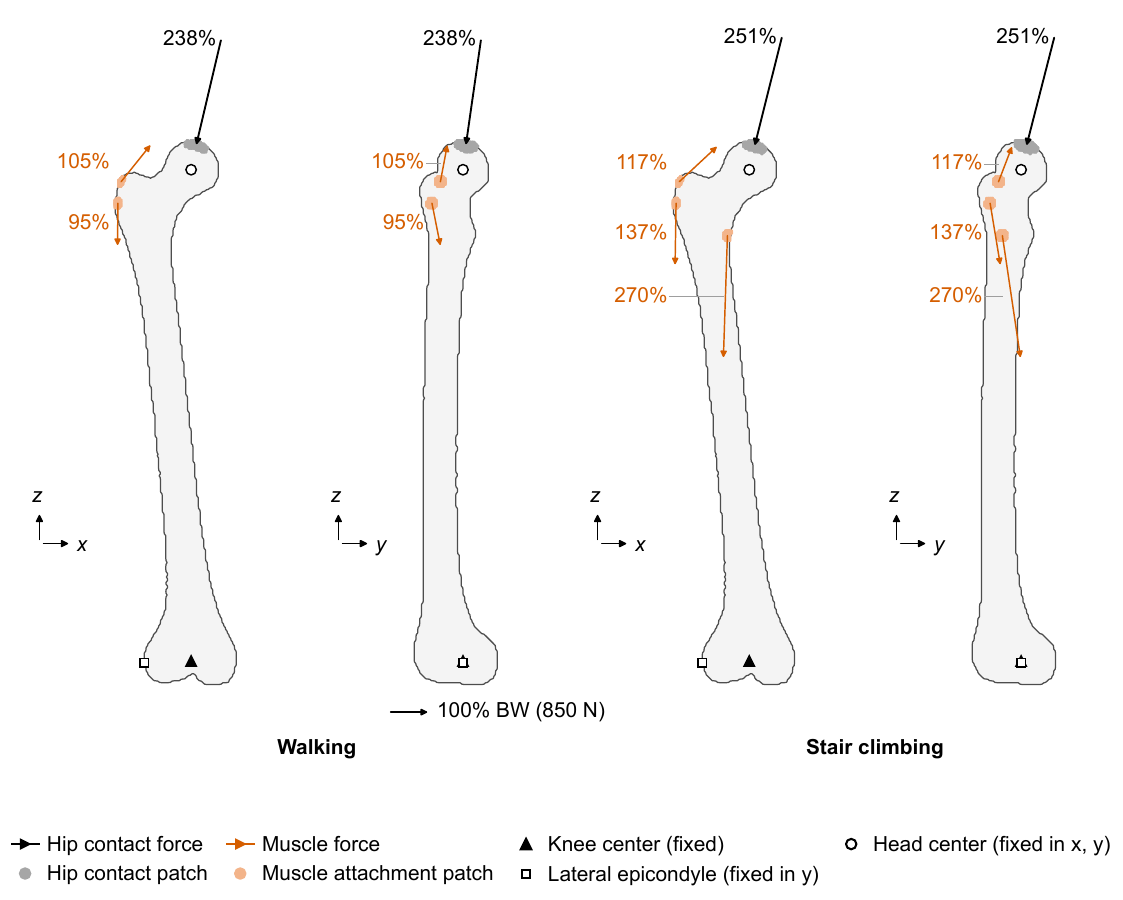}
    \caption{Front and side views of the loading conditions for the human femur model. (\emph{left}) Walking loads (\emph{right}) Stair-climbing loads. Forces are expressed relative to body weight -- modeled as 850N.}
    \label{fig:schematics}
\end{figure}

\subsection{Application: concurrent multiscale design for human femoral implants}

The two human femur applications use the same algorithm and numerical settings.
The two design problems are implants for a cortical mid-shaft and proximal cancellous defect.
The femur is discretized with 1.5mm Q1 elements, with walking and stair-climbing loads following \citet{Heller2005}, and boundary conditions following \citet{Speirs2007}.
Forces are normalized by a modeled body weight of 850N.
The design regions and modeling setup are visualized in Figure \ref{fig:schematics}.
The scaffold material is Ti-6Al-4V, with a Young's modulus of 110GPa.
All nine components of each class tensor are supplied by the continuum stage and enter the macroscale response directly.

Both optimizations begin with 12 flow-initialized cell classes, and both jointly update class latents and their spatial assignments following the concurrent process of Section \ref{sec:concurrent}.
Binary designs are periodically realized and verified, and the highest-porosity admissible iterate is retained.
The mechanical response is set to remain within 2\% of the intact reference under both loading conditions.

\subsubsection{Cortical mid-shaft defect}

A 30mm annular cortical defect is replaced with 777 blocks of 3mm lattice cells.
The objective is maximum porosity subject to whole-femur compliance constraints.
This objective both reduces the volume of implanted material and provides space for potential bone ingrowth, while maintaining the mechanical response of the intact femur.
This reflects the trade-off between osseointegration and structural stiffness in porous titanium implants \cite{Kelly2021,VanderStok2012}.
The physical feature-size floor is 234$\mu$m (10 voxels at 3mm).

\begin{figure}
    \centering
    \includegraphics[width=\linewidth]{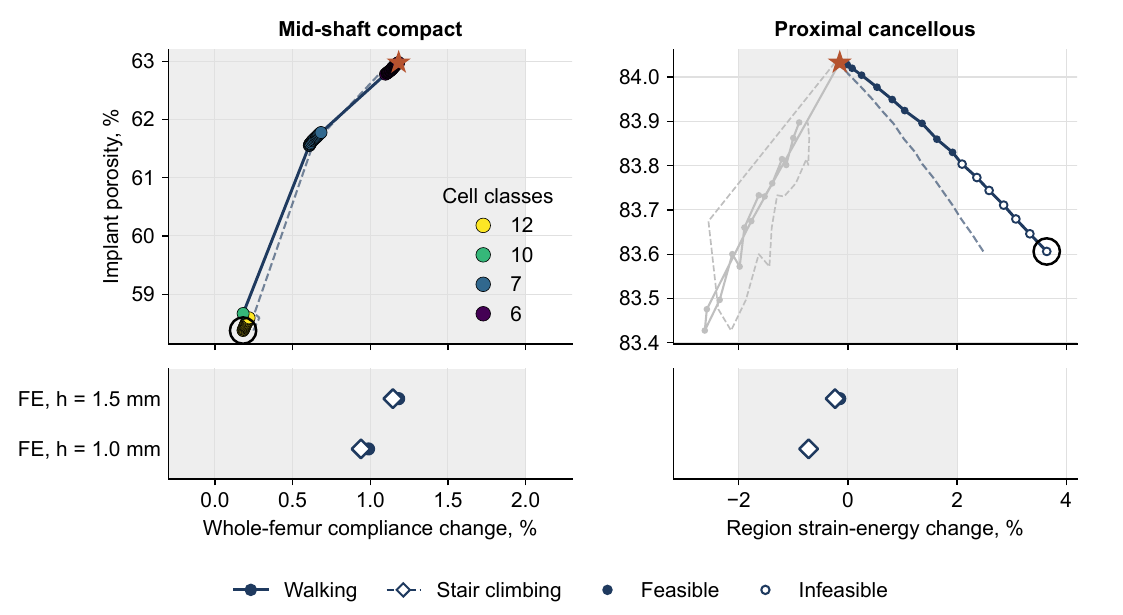}
    \caption{Concurrent optimization paths (\emph{top}) and verification of the reported designs (\emph{bottom}). Each path traces verified implant porosity against the constrained response under walking (solid) and stair-climbing (dashed) loads, from the initial design (ring) to the reported design (star); the shaded band is the $\pm2\%$ bound and gray segments follow the reported design. Mid-shaft points are colored by the number of active cell classes; the proximal design retains all 12 classes throughout. Bottom: finite-element verification of the reported designs on the 1.5mm design mesh and a refined 1.0mm mesh.}
    \label{fig:trajectories}
\end{figure}

The optimized scaffold achieves 62.98\% porosity compared with 58.38\% initially. 
Final verified compliance deviations are +1.18\% for walking and +1.15\% for stair climbing, both within the prescribed bounds.
Refinement of the macroscopic mesh from 1.5mm to 1.0mm gives corresponding deviations of +0.99\% and +0.94\% (Figure \ref{fig:trajectories}, \emph{bottom left}).

\begin{figure}
    \centering
    \includegraphics[width=\linewidth]{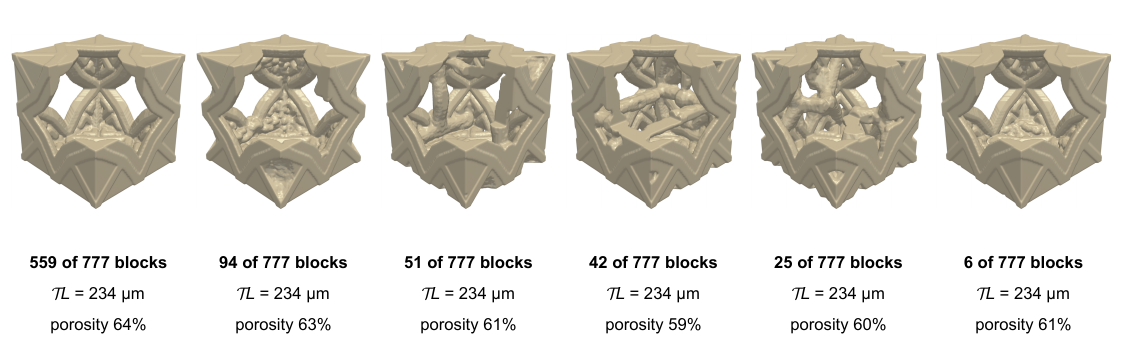}
    \caption{The six active cell classes in the optimized cortical mid-shaft scaffold, showing block populations, feature thickness, and porosity.}
    \label{fig:r2cells}
\end{figure}

Of the 12 available classes, only six remain active, shown in Figure \ref{fig:r2cells}, occupying 559, 94, 51, 42, 25 and 6 blocks.
The dominant class therefore occupies 72\% of the macroscale design blocks.
This is perhaps unsurprising due to the near-cylindrical symmetry of the implant.
Their cell porosities range from 59\% to 64\%, and all meet the 234$\mu$m thickness criterion.

The optimization path (Figure \ref{fig:trajectories}, \emph{left}) rises almost vertically: porosity increases in two marked steps, each coinciding with the merging of cell classes (from ten to seven, then to six), while compliance moves by only one percentage point. 
All 75 iterates remain admissible -- printable and within the response bound -- and the porosity-maximizing design occurs at iteration 73. 
All active cells reach the prescribed feature-size floor while compliance remains within the 2\% band, suggesting that feature size is the more restrictive condition in the obtained design.
Because only whole-femur compliance is constrained, the implant region carries 15--16\% more strain energy than the intact cortex it replaces.
This excess is present throughout the region: the element-wise deviation averages the same 15--16\%, as nearly all elements deviate in the same direction.

\begin{figure}
    \centering
    \includegraphics[width=\linewidth]{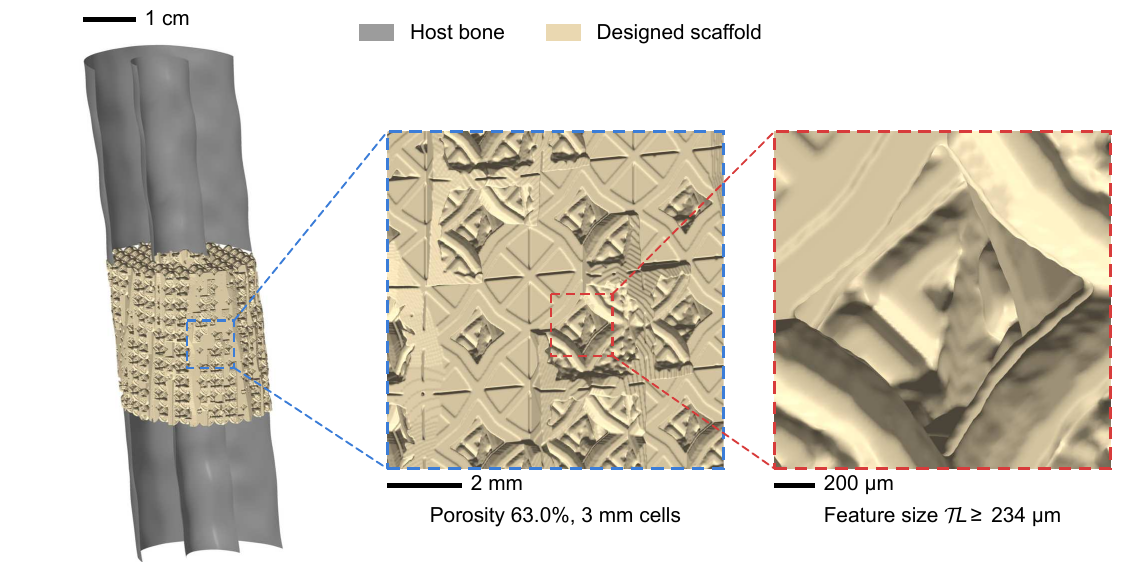}
    \caption{Multiscale visualization of the proposed human mid-shaft femoral scaffold.}
    \label{fig:r2implant}
\end{figure}

Figure \ref{fig:r2implant} shows the realized scaffold within the sectioned femur and at lattice-level (millimeters) and feature-level (micrometers) detail.
Optimization required 62 min on a single NVIDIA B200, followed by 101s for final verification.

\subsubsection{Proximal cancellous defect}

The second application replaces the femoral-neck and intertrochanteric cancellous core with 258 blocks of 6mm cells.
The feature-size threshold remains 234$\mu$m.
Because whole-femur compliance is relatively insensitive to this region, the mechanical constraint is imposed instead on implant-region strain energy.

The best verified scaffold achieves 84.03\% porosity.
Regional strain-energy deviations are $-0.15\%$ under walking loads, and $-0.24\%$ under stair-climbing loads.
Whole-femur compliance changes by only $-0.04\%$ and $-0.03\%$, respectively, although this is more reflective of the nature of the problem itself, rather than the proposed implant.

With like-for-like material assignments on a 1.0mm mesh, the regional strain-energy deviations become $-0.75\%$ and $-0.72\%$ (Figure \ref{fig:trajectories}, \emph{bottom right}).
The regional constraint is nevertheless sensitive to the discretization of the cortical-cancellous boundary, as resampling the bone model at 1.0mm changes the strain energy by 13--15\%.
The close response matching should therefore be interpreted within the adopted discretization.
Although the regional total is matched, element-wise strain-energy deviations average 10--12\%: deviations of opposite sign cancel within the region.

\begin{figure}
    \centering
    \includegraphics[width=\linewidth]{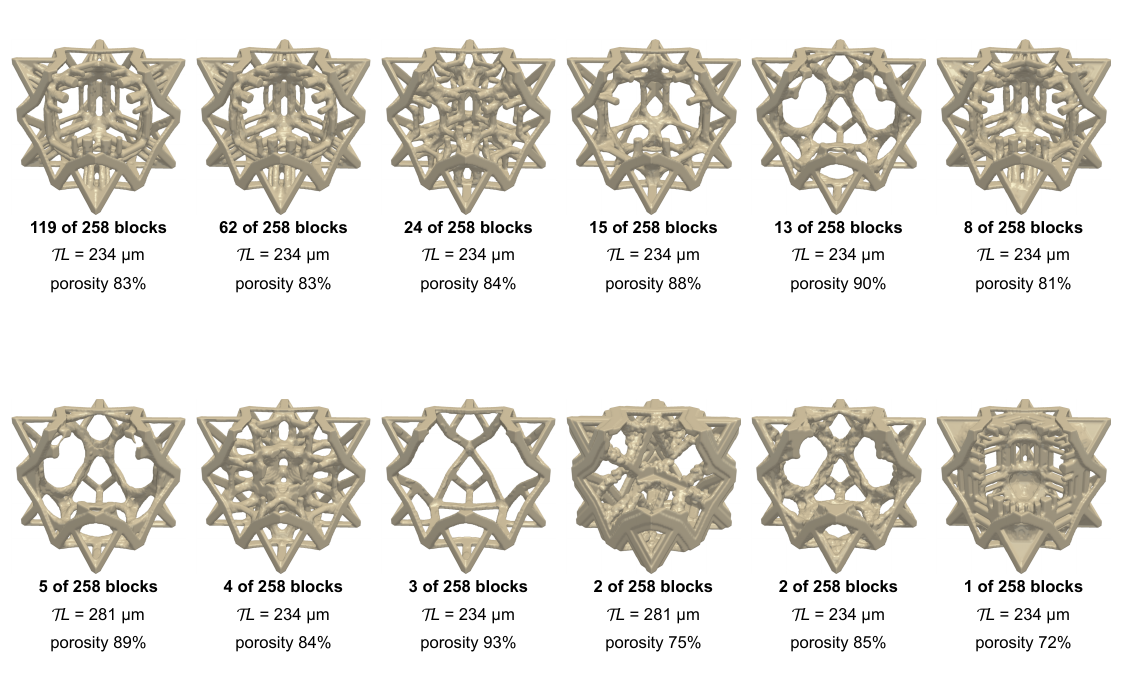}
    \caption{The twelve active cell classes in the optimized proximal cancellous scaffold, showing block populations, feature thickness, and porosity.}
    \label{fig:r3cells}
\end{figure}

Unlike the cortical case, all 12 classes remain active, and are illustrated in Figure \ref{fig:r3cells}.
Porosities range from 72\% to 93\%.
Block populations range from 119 to 1 -- considerably more spread than in the cortical design case.
It is notable that, once again, most cells satisfy the feature-size criterion at its floor, but two have a measured thickness of 281$\mu$m -- representing an extra voxel of thickness.

The optimization path (Figure \ref{fig:trajectories}, \emph{right}) begins outside the bound, with a +3.6\% regional strain-energy deviation at the first verified design, a consequence of the initialization. 
It then moves into the band while gaining porosity, reaching 84.03\% at the reported design. 
Subsequent iterates overshoot below $-2\%$ as the soft class assignment leads the realized design during temperature annealing; the trajectory recovers feasibility at lower porosity and is patience-stopped, so the best verified admissible iterate is reported.

\begin{figure}
    \centering
    \includegraphics[width=\linewidth]{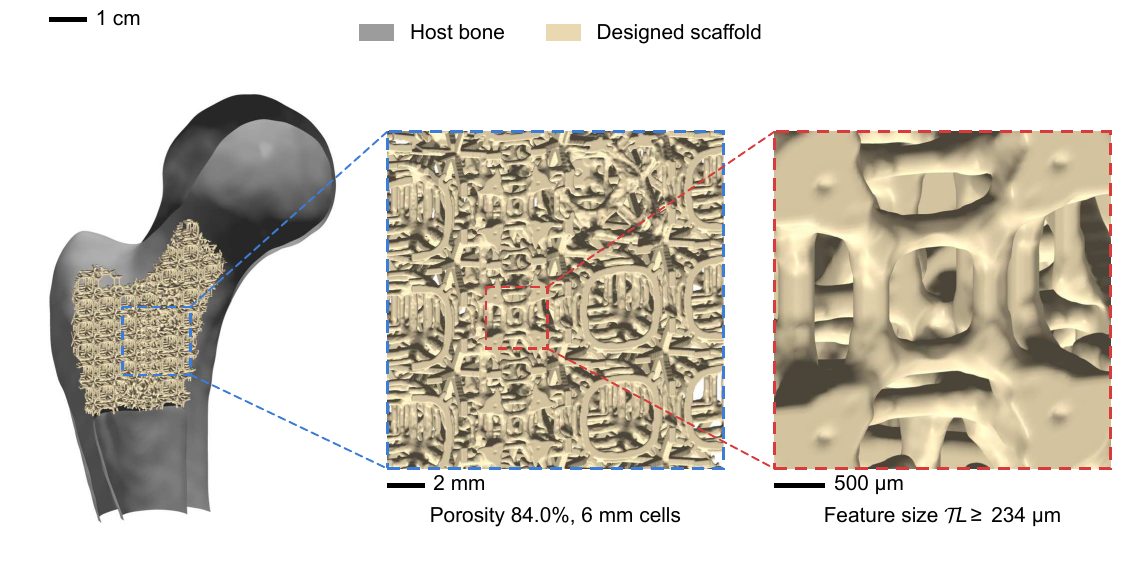}
    \caption{Multiscale visualization of the proposed  proximal cancellous femoral scaffold.}
    \label{fig:r3implant}
\end{figure}

The realized scaffold is shown in Figure \ref{fig:r3implant}. Optimization required 34.7 min, followed by 104s for final verification.

Neither concurrent formulation constrains the element-wise field: the mid-shaft design constrains whole-femur compliance and the proximal design the total regional strain energy. 
The resulting element-wise deviations average 15--16\% and 10--12\% (95th percentiles around 30\%), within the remodeling dead zone of strain-energy-based adaptation models, in which no net remodeling is predicted for local deviations of up to $\pm$35--75\% from the intact stimulus \cite{HUISKES1992}.
Constraining the local field directly, as in the cell-level case, can be an extension of the same adjoint formulation, at the cost of porosity, since element energies are already available from the adjoint solve.

\clearpage

\section{Discussion}\label{sec:discussion}

\begin{figure}
    \centering
    \includegraphics[width=\linewidth]{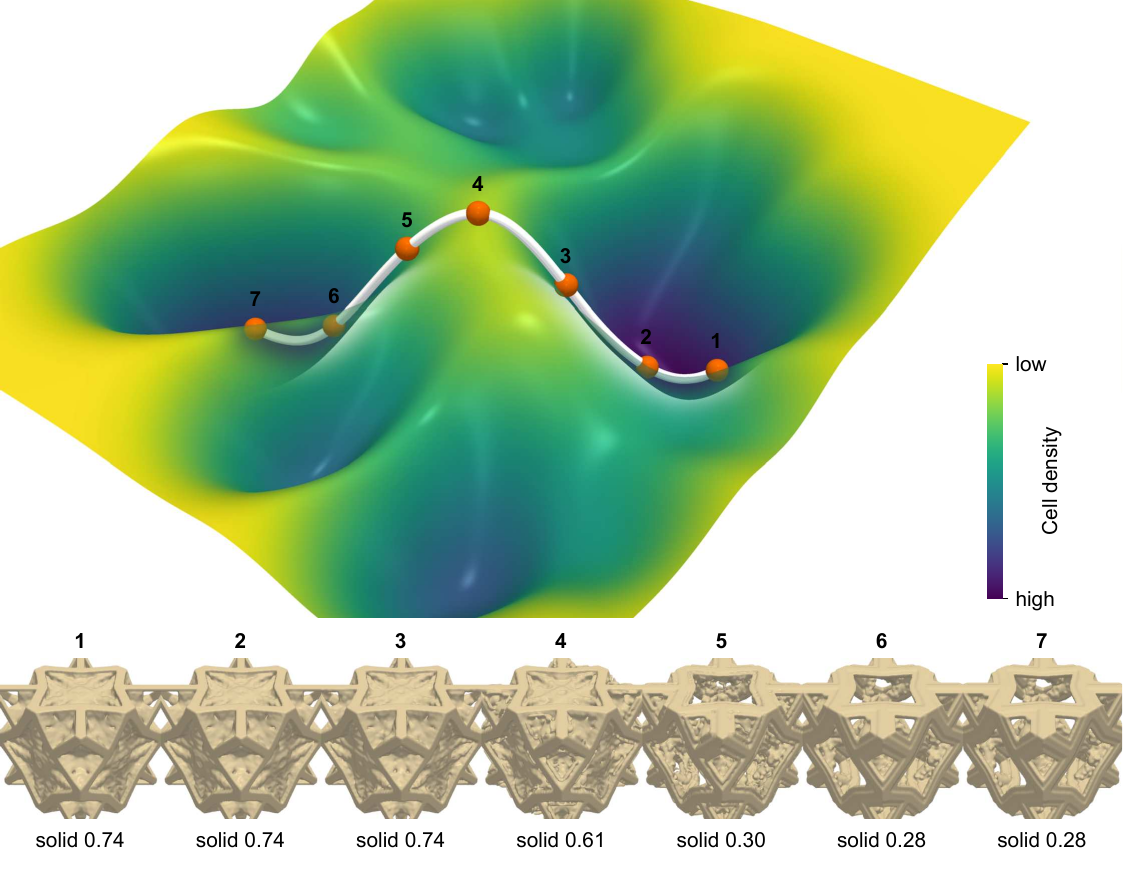}
    \caption{Organization of the learned latent space. (\emph{top}) density of 4,000 validation-cell latents in their two leading principal components, shown inverted so that densely populated regions form wells; no labels or properties are used. The tube follows a straight latent path between two flow-generated cells in the two most populated wells, with markers at the decoded frames below. (\emph{bottom}) Decoded cells along the path, with their solid fractions. The cell changes abruptly between two architectural families at the ridge rather than morphing gradually.}
    \label{fig:latentwalk}
\end{figure}

\begin{figure}
    \centering
    \includegraphics[width=\linewidth]{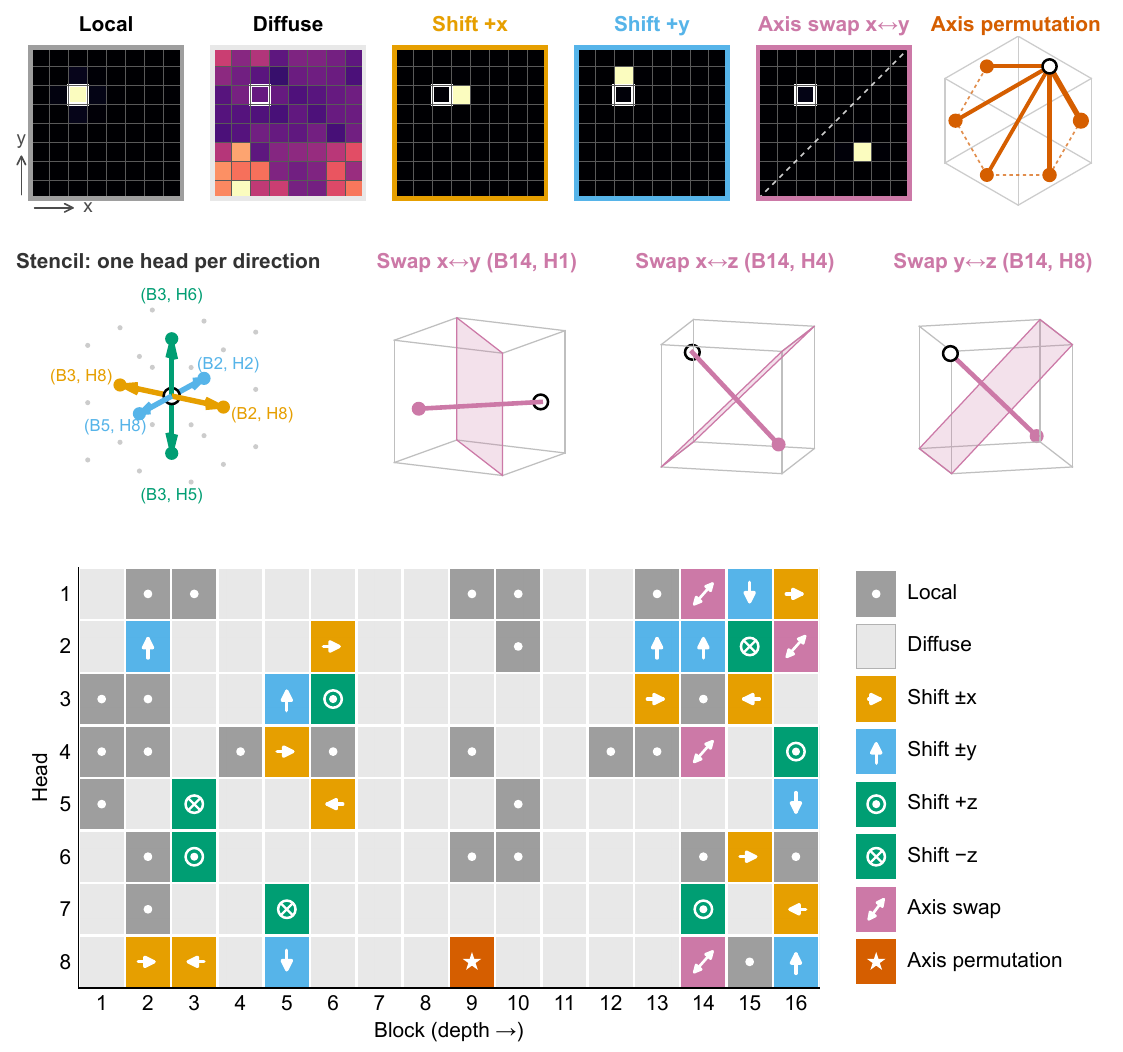}
    \caption{Attention patterns learned by the flow transformer. (\emph{top}) Attention of a single query token (white box) on an $8\times8$ slice of the $8^3$ token grid for representative heads, and, for the axis-permutation head, its attention to the five coordinate-permuted copies of the query, viewed along the cube diagonal. 
    (\emph{middle}) Three-dimensional detail of the six nearest-neighbor shift heads, each labeled (block, head), and the three axis-swap heads, each attending to the query's image under an exchange of two coordinate axes. 
    (\emph{bottom}) Classification of all 128 heads by block and head index.}
    \label{fig:attention}
\end{figure}

The results show that the proposed method reaches prescribed stiffness targets with printable cells where neither generation nor gradient-based refinement succeeds alone. 
This section examines the learned representations to explain why: how the latent space organizes candidate designs, and what spatial operations the generator has learned.

\paragraph{Design families and the role of the generative model.}

Projected onto their leading principal components, the validation latents form distinct wells, each corresponding to a family of cells sharing a common architecture (Figure \ref{fig:latentwalk}). 
These families emerge without supervision. 
Indeed, a straight path (in latent space) between cells in the two most populated wells does not produce a gradual blend. 
The decoded cell remains a closed-wall architecture ($v_f=0.74$) over the first half of the path and switches abruptly to a more open-frame architecture ($v_f=0.28$) at the ridge between wells. 
The same behavior occurs between other family pairs: 19 of 20 straight paths preserve topology over most of their length and then change abruptly.

The sharp transitions suggest that the decoder represents distinct architectural families rather than interpolating smoothly between lattices.
This feature also clarifies the division of labor in the proposed method. 
Gradient-based refinement is effective within a family, where small latent changes produce small, smooth changes in stiffness, but it cannot cross between families: refinement from random latents never reached a printable cell.
This defines the role of the generative model: supplying the family.
The rectified-flow places the search in the ``correct'' well, and exact refinement then closes the remaining gap. 
Neither component is sufficient alone. 
The best of 256 flow samples misses the rabbit target by 12.5\%, and refinement driven by a learned surrogate reduces the predicted error below 0.03\% while the exact error remains near 12\%. 
Only exact refinement from generated cells reaches the 1\% tolerance.

\paragraph{Symmetries and localities exhibited by the attention mechanism}

It is notable that the rectified-flow transformer receives no explicit spatial operators. 
Positions enter only through rotary positional embeddings.
Its attention heads, nevertheless, specialize into interpretable operations (Figure \ref{fig:attention}).
A fifth of the attention heads -- 25 -- act as nearest-neighbor shifts.
Each head attends almost exclusively to the adjacent token in one of the six axis directions.
Together, these heads form the six-point stencil of a finite-difference operator.
Three heads in block 14 attend to the query's image under an exchange of two coordinate axes (50\%, 93\% and 89\% of attention on that token), one head per pair of axes, and a single head in block 9 distributes its attention across all five coordinate-permuted images of the query (87\%). 

Head ablations demonstrate these roles to be genuinely functional.
If one zeroes the permutation head in block 9 during unconditional generation, the generated cells fragment -- the median number of disconnected components rises from 1 to 5. Zeroing four randomly selected diffuse heads leaves them intact. 
Removing the shift heads preserves the overall layout but produces granular surfaces with, on average, more than a thousand disconnected fragments per cell. 
In fact, these operations form early in training, being complete within 50,000 of about 175,000 steps, and the same heads retain the same roles at every later checkpoint, including all fine-tuning stages.
The rectified-flow transformer has thus learned spatial operations suited to periodic lattice cells -- local continuity and comparison across coordinate axes.

\paragraph{Limitations}

Several limitations qualify these results.
The cell-level validation targets a single transversely isotropic bone tensor whose anisotropy and Poisson ratios are modeling assumptions; broader validation against measured bone tensors is needed. 
For this demanding target the generator's samples occupy a single narrow design family, so success depends on exact refinement rather than on generative diversity. 
The concurrent formulations constrain whole-femur compliance or total regional strain energy rather than the element-wise field.
Local deviations therefore remain, although they lie within the remodeling dead zone assumed in strain-energy-based adaptation models \cite{HUISKES1992}.
A local constraint would be required to match the field more closely, albeit likely reducing the achievable porosity.
The regional strain-energy constraint is also sensitive to how the cortical-cancellous boundary is resolved: resampling the bone model at 1.0mm changes the intact regional energy by 13--15\%. 
The femur models use idealized walking and stair-climbing loads, simplified boundary conditions and linear-elastic homogenized bone properties, and homogenization assumes separation between cell and implant length scales, which is only approximate for 3--6mm cells. Graded baselines are independent re-implementations rather than the published pipelines, whose full lattice library is not available. 

\section{Conclusion}\label{sec:conclusion}

This work developed a concurrent two-scale topology optimization framework.
Lattice microstructures, initialized by a conditional rectified-flow model, were refined through differentiable finite-element homogenization in latent space.
At the scale of bone scaffolds, cell geometries and their spatial assignments are updated jointly.
Mechanical constraints were imposed upon the homogenized structure, and printability -- measured by feature thickness -- was assessed on realized cells.

The rabbit femur benchmark demonstrated the importance of combining conditional generation with exact refinement.
Neither flow-sampling alone, nor refinement of a random initialization achieved acceptable accuracy.
All eight flow-started runs, on the other hand, reached the prescribed six-modulus target within 1\% while meeting the feature size requirement.
A single repeated cell inverse-designed this way reproduced the intact femur's global compliance and regional strain energy more closely than any graded baseline design.

The human femur applications demonstrate the concurrent formulation under two different mechanical objectives.
Optimization of the cortical mid-shaft scaffold achieved a porosity of 63\%, while maintaining a whole-femur compliance within 2\% of the intact reference.
For a cancellous proximal scaffold, the method achieved 84\% porosity.
It is notable that these aggregate constraints do not ensure local response matching, and elementwise strain-energy deviations remain.
Further, the proximal result is sensitive to the discretization of the cortical--cancellous boundary.

The learned representations also help explain the method's behavior, in that distinct architectural families appear in the projected latent distribution.
The flow transformer's attention heads exhibit local continuity, in the form of learning a finite-difference-type stencil, and coordinate-symmetry operations.
These observations suggest that the generative model learns geometric structure useful for constructing coherent periodic cells.

The remaining limitations include the use of idealized bone properties and loading conditions, the sensitivity of regional responses to discretization, and the approximate separation of cell and implant length scales.
Within these limits, the results demonstrate that generative microstructure design can be coupled directly to numerical mechanics and concurrent structural optimization, in a way that can be made printability-aware.

\section{Acknowledgements}
The authors gratefully acknowledge discussions with Prof. Prahalada Rao (Virginia Tech) and Prof. Shaghayegh Bagheri (GMU) on topology optimization of implants. S.S. would like to acknowledge NSF CNS-2535293 fund for supporting this work.








\clearpage
\appendix
\section{Nomenclature}\label{app:A1}

\begingroup
\small
\setlength{\LTpre}{0pt}
\setlength{\LTpost}{0pt}
\renewcommand{\arraystretch}{1.15}

\begin{longtable}{@{}p{0.23\linewidth}p{0.71\linewidth}@{}}
\toprule
\textbf{Symbol} & \textbf{Description} \\
\midrule
\endfirsthead

\toprule
\textbf{Symbol} & \textbf{Description (continued)} \\
\midrule
\endhead

\bottomrule
\endfoot

\multicolumn{2}{@{}l}{\textit{Geometry and learned representations}} \\
$X,\tilde X$ &
Binary $64^3$ symmetry octant and decoded occupancy-probability field. \\

$\mathcal D,\mathcal R$ &
Frozen decoder and binary unit-cell realization operator. \\

$z,z_a$ &
Decoder-space latent and decoder-space latent of cell class $a$. \\

$z_0,z_\tau,z_1$ &
Gaussian initial, intermediate, and terminal standardized flow latents. \\

$\mu_z,\sigma_z$ &
Scalar mean and standard deviation used to standardize autoencoder latents for flow training. \\

$N_X,d_z$ &
Number of octant voxels and scalar latent dimension. \\

$y,\bar y,c$ &
Raw descriptor vector, standardized descriptors, and masked conditioning input. \\

$\mathcal M,\mathcal{\widetilde M}$ &
Component-level and group-level conditioning masks. \\

$\tau,v_\theta$ &
Flow pseudo-time and learned transport velocity field. \\

$\mathcal L$ &
Loss function, defined separately for autoencoder training, flow matching, and inverse design. \\

\addlinespace[4pt]
\multicolumn{2}{@{}l}{\textit{Geometric and material descriptors}} \\

$v_f,v_{f,a},v_b$ &
Cell solid fraction, solid fraction of class $a$, and relaxed block solid fraction. \\

$L,\mathcal T,T_{\mathrm{tol}}$ &
Physical unit-cell edge length, normalized fifth-percentile thickness, and physical thickness threshold. \\

$t(x),Q_{0.05}$ &
Local solid thickness and fifth-percentile operator. \\

$H_\beta,m_\xi$ &
Smooth Heaviside projection and relaxed stiffness coefficient of element $\xi$. \\

$\rho_{\min},\beta$ &
Relative void stiffness floor and projection sharpness parameter. \\

$\rho_{ij}$ &
Normalized coupling ratio between normal stiffness components $i$ and $j$. \\

$\tau_a,\tau_{\mathrm{tol}}$ &
Differentiable thin-feature fraction of class $a$ and its admissible upper bound. \\

$\phi$ &
Verified mean porosity of the assembled implant. \\

\addlinespace[4pt]
\multicolumn{2}{@{}l}{\textit{Homogenization and structural mechanics}} \\

$C^H,c^H,S^H$ &
Homogenized elasticity, its dimensional nine-component orthotropic stiffness vector, and Voigt compliance matrix. \\

$C_b,C_\xi$ &
Implant-scaled block constitutive tensor and macroscopic element constitutive tensor. \\

$D^{(0)},D_b,D_\xi$ &
Dimensionless reference-solid elasticity tensor and Voigt constitutive matrices for block $b$ and element $\xi$. \\

$E_0,E_s,\nu_s$ &
Reference solid modulus, implant solid modulus, and reference-solid Poisson ratio. \\

$e,\hat e,e^\star$ &
Engineering-modulus vector, computed engineering moduli, and prescribed target moduli. \\

$\epsilon_{\max},\epsilon_{\mathrm{tol}}$ &
Maximum relative engineering-modulus error and acceptance tolerance. \\

$\Omega_{\mathrm{cell}},\Omega_{\mathrm{cell},\xi}$ &
Periodic unit-cell domain and its element subdomain $\xi$. \\

$\Omega,\Omega_d,\Omega_b$ &
Macroscopic domain, implant replacement region, and design-block domain. \\

$A^{(i)},\chi^{(i)},\varepsilon^{(i)}$ &
Imposed strain, periodic displacement fluctuation, and strain field in cell homogenization. \\

$u^{(i)},f^{(i)},\lambda^{(i)}$ &
Macroscopic displacement, external load vector, and adjoint displacement for load case $i$. \\

$K,K_d$ &
Assembled stiffness matrix and implant-region stiffness contribution. \\

$J^{(i)},J_0^{(i)},r^{(i)}$ &
Mechanical response (compliance or regional strain energy), intact reference response, and relative deviation. \\

$x_{b,k}$ &
Physical constitutive coordinate $k$ of design block $b$. \\

\addlinespace[4pt]
\multicolumn{2}{@{}l}{\textit{Concurrent optimization}} \\

$N_b,N_c,N_L$ &
Numbers of design blocks, unit-cell classes, and macroscopic load cases. \\

$\alpha_{ba},w_{ba},T$ &
Class-assignment logit, soft assignment weight, and softmax temperature. \\

$a(b)$ &
Discrete class assigned to design block $b$ during verification. \\

$\delta$ &
Admissible magnitude of relative mechanical-response deviation. \\

\end{longtable}
\endgroup

\noindent\textit{Index conventions.}
The index $\xi$ denotes a voxel or finite element, $a$ a unit-cell
class, $b$ a design block, and $k$ a constitutive component.
Spatial indices $p,q,r,s\in\{1,2,3\}$ follow the Einstein
summation convention.
In cell homogenization, $i,j\in\{1,\ldots,6\}$ index Voigt
strain cases; in macroscale analysis, $i$ indexes load cases.
\clearpage

\printcredits

\bibliographystyle{unsrtnat}

\bibliography{cas-refs}



\end{document}